\documentclass[a4paper,11pt]{article}
\pdfoutput=1 

\usepackage{jheppub}
\usepackage[T1]{fontenc} 

\title{MSSM-inspired multifield inflation}

\author[a]{M.~N.~Dubinin,}
\author[a,b]{E.~Yu.~Petrova,}
\author[a,1]{E.~O.~Pozdeeva\note{Corresponding author.},}
\author[b]{M.~V.~Sumin,}
\author[a]{and S.~Yu.~Vernov}

\affiliation[a]{Skobeltsyn Institute of Nuclear Physics, Lomonosov Moscow State University,\\Leninskiye Gory~1, 119991, Moscow, Russian Federation}
\affiliation[b]{Physics Department, Lomonosov Moscow State University,\\Leninskiye Gory~1, 119991, Moscow, Russian Federation}

\emailAdd{dubinin@theory.sinp.msu.ru}
\emailAdd{petrova@theory.sinp.msu.ru}
\emailAdd{pozdeeva@www-hep.sinp.msu.ru}
\emailAdd{mv.sumin@physics.msu.ru}
\emailAdd{svernov@theory.sinp.msu.ru}

\abstract{Despite the fact that experimentally with a high degree of statistical significance only a single Standard Model--like Higgs boson
is discovered at the LHC, extended Higgs sectors with multiple scalar fields not excluded by combined fits of the data are more
preferable theoretically for internally consistent realistic models of particle physics. We analyze the inflationary scenarios which could be induced by the two-Higgs-doublet potential of the Minimal Supersymmetric Standard Model (MSSM) where five scalar fields have non-minimal couplings to gravity. Observables following from such MSSM-inspired multifield inflation are calculated and a number of consistent inflationary scenarios are constructed. Cosmological evolution with different initial conditions for the multifield system leads to consequences fully compatible with observational data on the spectral index and the tensor-to-scalar ratio. It is demonstrated that the strong coupling approximation is precise enough to describe such inflationary scenarios.}


\arxivnumber{1705.09624 [hep-ph]}

\begin{document}
\maketitle
\flushbottom

\section{Introduction}
\label{sec:intro}

The inflationary models, which solve successfully the horizon, flatness and relic problems \cite{general1_0, general1_1, general1_2} and generate the primordial density perturbations finally initiating the formation of galaxies and large-scale structure \cite{general2_0, general2_1, general2_2}, are the most reasonable models for the evolution of the early Universe. In simplest case inflation is controlled by a single scalar field (the inflaton) with an effective potential that plays a role of the cosmological constant during inflation.

A fundamental step towards the unification of physics at all energy scales could be the possibility to describe the inflation using particle physics models. In numerous models (for a review see~\cite{Lyth:1998xn}) the role of the inflaton has been performed by the Standard Model (SM) Higgs boson~\cite{Cervantes-Cota1995, higgsinf_0, higgsinf_1, higgsinf_2, higgsinf_3, higssinflRG_0, higssinflRG_1, higssinflRG_2} or a boson in Grand Unified Theories (GUTs)~\cite{GUT_Inflation_0, GUT_Inflation_1} or a scalar boson in supersymmetric (SUSY) models~\cite{SUSEinflation_0, SUSEinflation_1, SUSEinflation_2} (see~\cite{MazumdarRev,Ferrara:2015cwa} as reviews). A number of advantages of simplified SUSY GUTs in comparison with nonsupersymmetric GUTs such as naturally longer period of exponential expansion and better stability of the effective Higgs potential with respect to radiative corrections due to cancelation of loop diagrams have been noted quite long ago~\cite{ellis}.

Thus, the implementation of inflationary scenario within a well-defined model of particle physics consistent with collider phenomenology where the inflaton is unambiguously identified is a longstanding problem. The only candidate on the role of the inflaton in the SM is the Higgs boson. The Higgs-driven inflation
\cite{higgsinf_0, higgsinf_1, higgsinf_2, higgsinf_3, higssinflRG_0, higssinflRG_1, higssinflRG_2}
was originally proposed as a single-field model based on the SM in the unitary gauge. This minimal model uses the Higgs isodoublet $\Phi$ interaction with gravity of the form $\xi R \Phi^\dagger \Phi$ ($R$ is the scalar curvature and $\xi$ is a positive constant). The Higgs-driven inflation leads to the spectral index value $n_\mathrm{s}=0.967$ and the tensor-to-scalar ratio $r=3\cdot 10^{-3}$ which are in agreement with the Planck Collaboration data~\cite{planck_0, planck_1,  PlanckIfl_0, PlanckIfl_1}.
However, the effects of Goldstone bosons should be included at an energy scale relevant to inflation in the model which is actually multifield.

For the Higgs-driven inflation it was found~\cite{KaiserHiggs} that the multifield effects are negligibly small during inflation and do not influence the observable quantities, such as the spectral index of primordial perturbations and the ratio of squared amplitudes for the tensor and the scalar perturbations (tensor-to-scalar ratio).
This fact and the known considerations about the need to extend the SM, which could be an effective limit of GUTs, supersymmetry, supergravity or other beyond the SM theories, lead to the belief that inflation that is compatible with recent observations~\cite{planck_0, planck_1, PlanckIfl_0, PlanckIfl_1} might have been generated by several fields. It has been shown~\cite{Kaiser:2013sna,Schutz:2013fua} that there is a class of inflationary models with two scalar fields non-minimally coupled to gravity that provides good agreement with the Planck data.   Hybrid inflation proposed in~\cite{hybrid_0, hybrid_1, hybrid_2} which involves the potential of two scalar fields ensures inflationary expansion, explains the observed spectrum of density fluctuations not requiring the unnatural scalar field amplitudes at the Planck scale. At the same time, a greater degree of uncertainty arises in the theory. Identification of the two scalar fields in the framework of a gauge theory model is not simplified, the initial conditions are not unambiguously fixed~\cite{initial} theoretically in the two-dimensional field space and their tuning is needed to ensure adequate phenomenological consequences.

It should be noted that some tension is observed between the experimental data and predictions of the minimal model.  In to order to explain cosmic microwave background observables in the Higgs-driven inflationary scenario the parameter of non-minimal coupling should be very large ($\xi\sim 10^4$). So large value of $\xi$ is not satisfactory from general theoretical backgrounds because it leads to violation of perturbative unitarity at the scale $M_{Pl}/\xi$ which is smaller than the expected inflationary range above the $M_{Pl}/\sqrt{\xi}$ ($M_{Pl}$ denotes the Planck mass). In order to restore unitarity above the scale $M_{Pl}/\xi$,  "new physics" (new particles interacting with the SM ones) should be introduced which modify the SM Higgs potential. The more serious problem of a large $\xi$ value in the SM is the renormalization group evolution (RGE) of meaningful parameters which demonstrates unsa\-tis\-fac\-tory matching with the measured Higgs boson and top quark masses \cite{higgstop} as soon as the inflationary range of the order of $M_{Pl}/\sqrt{\xi}$ or above is concerned. In order to reduce the value of $\xi$, an extremely small value of the effective quartic coupling $\lambda_{eff}(\mu)$ near the Planck scale is needed. At the Higgs boson mass $m_h=125 \,GeV$ such value of $\lambda_{eff}(\mu)$ can be achieved at the top quark mass which is more than 2$\sigma$ below its observable central value \cite{small_lambda_0, small_lambda_1, small_lambda_2}. In this case, the value of $\xi$ necessary for a satisfactory inflationary scenario decreases thus allowing to avoid the problem of perturbative unitarity violation below the inflationary scale. Note that the GUT motivated inflationary model~\cite{Barvinsky:1994hx} predicts the same order of the parameter for a non-minimal coupling. However, there are cosmological models (see, for example~\cite{Tronconi:2017wps}) with the same function of the non-minimal coupling and even polynomial potential of the fourth order that could provide a suitable inflationary parameters at small values of~$\xi$.

Apparent tensions arising in connection with parameter matching of the Higgs-driven inflation model increase the popularity of models with new physics at the TeV and multi-TeV scales. New particles consistent with restrictions on the new physics imposed by the LHC data provide extensive opportunities to improve significantly the Higgs-driven inflationary model. Analogue of this single-field model for the multifield scenarios is based on an observation that redefined fields in the Einstein frame  practically coincide with primary fields in the Jordan frame at the low energy scale of the order of superpartners mass scale $M_{SUSY}$, reproducing the MSSM potential, while at the scale higher than the GUT scale the potential in the redefined fields can be slowly changing respecting the slow-roll approximation of an inflationary scenarios. This observation is sufficiently general.

Recent analyses \cite{recentattempts_0, recentattempts_1, recentattempts_2, recentattempts_3, recentattempts_4, recentattempts_5, Kaiser:2010yu, Kaiser} of multifield models showed that unlike the single-field models they generically provide density (entropy) perturbations which can induce the curvature perturbation to evolve beyond the cosmological horizon in the process of inflation~\cite{c_perturbations}. Evolution of density perturbations in multifield models should be studied in order to analyze new features in the observables such as non-Gaussianities~\cite{non_g} which are absent in the single-field inflationary models. Deviations of observable power spectrum calculated in multifield models from predictions of the single-field models could take place in the power of one of the three criteria~\cite{Kaiser,criteria}:
(i) noncanonical kinetic terms; (ii) violation of slow-roll approximation; (iii) nonstandard initial ground state (different from Bunch--Davies vacuum). The main feature of the multifield models which leads to nonstandard primordial spectrum is the ability of trajectories of slow-roll fields to rotate in the field space, that occurs due to the presence of bumps and ridges in the effective multifield potential. When the slow-roll field trajectories turn in the field space, nonstandard contributions to primordial spectra can be amplified enough to be detectable in the microwave background~\cite{Peterson:2010np, nonstandard_amplified_0, nonstandard_amplified_1}.

In this paper we analyze a multifield extension of the standard Higgs-driven inflation inspired by the MSSM.
A few general observations let us make first. In the framework of a sypersymmetric model the natural class of cosmological models are those with local supersymmetry (supergravity models). For the case when interactions at an energy scale below $M_{Pl}$ are described by an effective $N=1$  supergravity, the general form for the effective potential of scalar fields in the Einstein frame was derived in \cite{cremmer}. In the notation of \cite{nilles} the Lagrangian can be written as
\begin{equation}
\label{eq1}
L_B=e^{-G} \, \left[G_k (G^{-1})^k_i G^i +3\right]-\frac{1}{2}{\hat g}^2 Re \, f^{-1}_{\alpha \beta}
(G^i T^{\alpha j}_i z_j) (G^k T^{\beta j}_k z_j) + G^i_j D_\mu z_i D^\mu z^{*j} - \frac{1}{2} R
\end{equation}
where $G_k=\partial G/\partial \varphi^k$, ${\hat g}$ is the gauge coupling constant, $T$ are generators of the groups, $D_\mu$ are covariant with respect to gravity and gauge group, ($z_i$, $\chi_i$) is the chiral supermultiplet.

The K\"{a}hler potential  $G$ can be written in terms of the function $\phi$ which transforms as a real vector superfield and the superpotential $g_s$ in the following form:
\begin{equation}
G=3\log(-\phi/3)-\log(|g_s|^2).
\end{equation}

Equation~(\ref{eq1}) is a consequence of a Lagrangian written in terms of chiral superfields~$\tilde{\Phi}$:
\begin{equation}
\label{einhorn_jones}
L= {}-6 \int \, d^2 \theta \, {\cal E} \, \left[R-\frac{1}{4} ({\overline{\cal D}}^2-8 \, R) \tilde{\Phi}^\dagger\tilde{\Phi}+g_s \right] + h.c.,
\end{equation}
here $R$ is the superspace curvature and ${\cal E}$ is chiral density connected with local superspace basis (see \cite{bagger}).
One can show~\cite{einhorn} that minimal coupling to gravity which takes place for $\phi=z^*_i z_i - 3$ in the K\"{a}hler potential can be modified to a non-minimal coupling $R\to R+p(\tilde{\Phi})R$ instead of the first term in eq.~(\ref{einhorn_jones}) by the replacement $ \phi=z^*_i z_i - 3 -3 (p(z)+h.c.)/2$ for a given polynomial form $p$.\footnote{Taking frequent in the literature point of view that the main qualitative features at the scale of the order of $M_{Pl}$ are valid despite the loop effects of gravity, there is an opinion that in a simplest case for a single field any polynomial form $p(x)$ can be adjusted by taking $dg_s/dx=\sqrt{\left(3+p(x)\exp(G(x))\right)/2}$ (where $x$ is $Re \, z$, ($z_i$, $\chi_i$) is the chiral supermultiplet) with the following extension of a solution in the form of series expansion to complex $z$ \cite{goncharov, ellis:1983}.}

In the case of the MSSM natural choice is $p=\xi \bar \Phi_1 \bar \Phi_2$, where $\bar \Phi_1$ and $\bar \Phi_2$ are chiral Higgs-Higgsino superfields. This choice of $p$ in combination with the general form for the superpotential $g_s=\Lambda+\mu \bar \Phi_1 \bar \Phi_2$ ($\Lambda$ and $\mu$ are real constants) leads to problems of achieving a suitable inflationary scenario, see~\cite{einhorn,Ferrara:2010yw}. For $\xi$ parameter large enough when different regimes for the 'flat direction' $\tan \beta$ parameter are taken,  either there is no slow roll or the potential takes negative values. Not referring here to the possibility of the MSSM extension with a gauge singlet (non-minimal MSSM or NMSSM) where unsuitable behavior can be cured, we introduce non-minimal couplings in the non-holomorphic form
$(\xi_1 H^\dagger_1 H_1+\xi_2 H^\dagger_2 H_2)\, R$ (here $H_1$ and $H_2$ are $SU(2)$ spinors and $R$ is the Ricci scalar) that have no counterparts in supergravity. So only small electroweak quartic couplings $g_1$ and $g_2$ in the $D$-terms of eq.~(\ref{eq1}) which then appear in the tree-level scalar potential at the SUSY scale provide grounds to speak about 'MSSM-inspired' inflationary scenarios. This sort of model is not a direct extension of models associated with the MSSM which include scalar fields minimally coupled to gravity~\cite{Allahverdi:2010zp, Chatterjee:2011qr, Ibanez:2014swa}. The inflaton fields are identified as Higgs sector fields, thus, in this case one is talking about the multifield extension of the SM single-field Higgs inflation. Note that other realizations of the inflationary scenario in the MSSM are possible, when the inflaton is a combination of squark and slepton fields \cite{enqvist}, while the process of inflation is controlled by flat directions of the MSSM potential which are lifted by non-renormalizeable superpotential terms and soft supersymmetry breaking terms. It is assumed that $D$-terms in eq.~(\ref{eq1}) vanish in the hidden sector. A number of other options of the MSSM-inspired inflation can be found in \cite{SUSEinflation_0, SUSEinflation_1, SUSEinflation_2, MazumdarRev, Ferrara:2015cwa,ellis, Allahverdi:2010zp, Chatterjee:2011qr, Ibanez:2014swa, Antoniadis:2016aal}.

The model which is considered in the following sections includes two Higgs doublets coupled with gravity non-minimally. We focus on the two-Higgs doublet MSSM potential in the mass basis of scalar fields that has been analyzed starting from 1975~\cite{full_mssm}. This potential includes three massless Goldstone bosons and five massive Higgs bosons. Working in the physical gauge, in this paper we do not take Goldstone bosons into account and consider inflationary scenarios that include Higgs bosons only. We show that inflationary scenarios with suitable parameters $n_s$ and $r$ are possible at the scale corresponding to the Hubble parameter $H\sim 10^{-5}M_{Pl}$. By this way a MSSM-inspired extension of the original Higgs-driven inflation is constructed.

The structure of the paper is as follows. In section \ref{sec:2} we define the MSSM two-Higgs-doublet potential in the basis of mass eigenstates for the five Higgs bosons at the superparticle mass scale. The mixing angles of the $SU(2)$ field eigenstates are chosen in the form which is acceptable for the low-energy Higgs phenomenology. In section \ref{sec:3} the MSSM-inspired potential taken in the Jordan frame with the polynomial form of the non-minimal coupling function is transformed to the Einstein frame. Equations of motion in the Friedmann--Lemaitre--Robertson--Wal\-ker (FLRW) metric are described in section \ref{sec:4}. Numerical integration of the equations of motion with the initial conditions which are adjusted in a way suitable for reproduction of the observable values for the spectral index $n_\mathrm{s}$ and the tensor-to-scalar ratio $r$ is preformed in section \ref{sec:5}. In section \ref{sec:6} we discuss briefly the strong coupling (SC) approximation for the MSSM-inspired potential under consideration. Results are summarized in section~\ref{sum}.


\section{The MSSM-inspired Higgs potential}\label{sec:2}

Two Higgs doublets of the MSSM can be parameterized using the $SU(2)$ states
\begin{eqnarray}
\label{dublets}
\Phi_1 = \left(\begin{array}{c} -i \omega_1^+ \\ \frac{1}{\sqrt{2}} (v_1+\eta_1+i \chi_1) \end{array} \right),\\
\Phi_2 = \left(\begin{array}{c} -i \omega_2^+ \\ \frac{1}{\sqrt{2}} (v_2 +\eta_2+i \chi_2) \end{array} \right),
\end{eqnarray}
where $\omega^+_{1,2}$ are complex scalar fields, $\eta_{1,2}$ and $\chi_{1,2}$ are real fields, the vacuum expectation values $v_{1}$ and $v_2$ are usually redefined in ($v$, $\tan \beta$) parametrization: $v=\sqrt{v^2_1+v^2_2}$ and $\tan \beta=v_2/v_1$ ($v=246\, GeV$). Two doublets $\Phi_{1}$ and $\Phi_{2}$ can be used to form the $SU(2) \times U(1)$ invariant and renormalizable effective potential which breaks gauge symmetry.

The most general two-doublet effective potential can be written as~\cite{hh93}:
\begin{eqnarray}
\label{genU}
V(\Phi_1,\Phi_2) &=&
- \, \mu_1^2 (\Phi_1^\dagger\Phi_1) - \, \mu_2^2 (\Phi_2^\dagger
\Phi_2) - [ \mu_{12}^2 (\Phi_1^\dagger \Phi_2) +h.c.] \nonumber \\
&+& \lambda_1
(\Phi_1^\dagger \Phi_1)^2
      +\lambda_2 (\Phi_2^\dagger \Phi_2)^2
+ \lambda_3 (\Phi_1^\dagger \Phi_1)(\Phi_2^\dagger \Phi_2) +
\lambda_4 (\Phi_1^\dagger \Phi_2)(\Phi_2^\dagger \Phi_1) \nonumber \\
&+& \left[\frac{\lambda_5}{2}
       (\Phi_1^\dagger \Phi_2)(\Phi_1^\dagger\Phi_2)+ \lambda_6
(\Phi^\dagger_1 \Phi_1)(\Phi^\dagger_1 \Phi_2)+\lambda_7 (\Phi^\dagger_2 \Phi_2)(\Phi^\dagger_1 \Phi_2)+h.c.\right].
\end{eqnarray}

Let us consider the action in the Jordan frame
\begin{equation}
\label{action}
    S=\int d^4x\sqrt{-\tilde{g}}[f(\Phi_1,\Phi_2)\tilde{R}-\delta^{ab}\tilde{g}^{\mu\nu}\partial_\mu\Phi_a^{\dagger}\partial_\nu\Phi_b
    -\, V(\Phi_1,\Phi_2)],
\end{equation}
where $\tilde{g}$ is the determinant of metric tensor $\tilde{g}_{\mu\nu}$, and $R$ is the scalar curvature. The factor in front of the kinetic term is not dependent on fields, so the case of Brans--Dicke gravity-like models are beyond our analysis. However, $\delta^{ab}$ in front of the kinetic term is not narrowing the generality of consideration, see details in appendix \ref{apA}.
Variation of action with respect to metric tensor $\tilde{g}^{\mu\nu}$ and isodoublets $\Phi_a$ of the fields leads to the following equations
\begin{eqnarray}
f(\Phi_1,\Phi_2)\left[\tilde{R}_{\mu\nu}-\frac{\tilde{R}}{2}\tilde{g}_{\mu\nu}\right] = \left(\nabla_\mu\nabla_\nu - \tilde{g}_{\mu\nu}\nabla^\alpha\nabla_\alpha\right)f(\Phi_1,\Phi_2) \nonumber \\
+2\delta^{ab}\left[\partial_{\mu}(\Phi_a)^\dag{}\partial_{\nu}\Phi_b-\frac{1}{2} \tilde{g}_{\mu\nu}\partial_{\alpha}(\Phi_a)^\dag{}\partial^{\alpha}\Phi_b\right] - \frac12V(\Phi_1,\Phi_2)\tilde{g}_{\mu\nu},
\label{General_metric_equations}
\end{eqnarray}
\begin{equation}
2 \, \Box{}\Phi_a = -\frac{\partial{}f(\Phi_1,\Phi_2)}{\partial{}\Phi_a^\dagger}\tilde{R} + \frac{\partial{}V(\Phi_1,\Phi_2)}{\partial{}\Phi_a^\dagger}\vspace{0mm}, \label{General_fields_equations}
\end{equation}
where $a=1,2$,  $\nabla_\mu$ is a covariant derivative and the d'Alembert operator acting on the scalar fields is denoted by $\Box \equiv \frac{1}{\sqrt{-\tilde{g}}}\partial_\mu\left(\sqrt{-\tilde{g}}\tilde{g}^{\mu\nu}\partial_\nu\right)$.
In the following we are using notations and normalization conventions for the potential $V(\Phi_1,\Phi_2)$ in the generic basis (with $\lambda_{6,7}$ terms) from~\cite{Akhmetzyanova:2004cy_0, Akhmetzyanova:2004cy_1}, where the mass eigenstates for scalars were constructed.

Note that the potential in eq.~(\ref{genU}) explicitly violates CP invariance if parameters $\mu_{12}$, $\lambda_5$, $ \lambda_6$, or $ \lambda_7$ are complex-valued. For simplicity, we are not considering such possibility in the following. At the tree-level $\lambda_5$, $ \lambda_6$, and  $ \lambda_7$ are equal to zero in the MSSM two-doublet potential. Nonzero parameters $\lambda_{5,6,7}$ of the effective Higgs potential can be generated by radiative corrections coming from the sector of soft supersymmetry breaking terms, where scalars couple to quark superpartners. To simplify the analysis we will not consider this possibility remaining with the tree-level potential at the $M_{SUSY}$ scale. It is well-known that radiative corrections are large and in the context of this simplification (when the upper limit of the light CP-even state mass $m_h$ does not exceed the $Z$-boson mass $m_Z=91.2 \,GeV$) it is impossible to describe adequately the spectrum of Higgs boson masses. However, precision fitting of the collider data is not the primary purpose at this stage of consideration.

The mass basis of scalars is constructed in a standard way. The $SU(2)$ eigenstates ($\omega^\pm_a,\eta_a$ and $\chi_a$, $a=1,2$) are expressed through mass eigenstates of the Higgs bosons $h$, $H_0$, {$A$} and $H^\pm$ and the Goldstone bosons $G^0$, $G^\pm$  by means of two orthogonal rotations
\begin{eqnarray}
\left( \begin{array}{c} \eta_1 \\ \eta_2 \end{array} \right)={\cal O}_\alpha
\left( \begin{array}{c} H_0 \\ h  \end{array} \right), \qquad
\left( \begin{array}{c} \chi_1 \\ \chi_2 \end{array} \right)={\cal O}_\beta
\left( \begin{array}{c} G^0 \\ A  \end{array} \right), \qquad
\left( \begin{array}{c} \omega_1^\pm \\ \omega_2^\pm \end{array} \right)={\cal O}_\beta
\left( \begin{array}{c} G^\pm \\ H^\pm  \end{array} \right),
\end{eqnarray}
where the rotation matrix
\begin{equation}
{\cal O}_X=\left( \begin{array}{cc} \cos X & -\sin X \\ \sin X & \cos X  \end{array} \right), \qquad
X=\alpha, \beta.
\end{equation}
Masses of the CP-even scalars $h$ and $H_0$ are $m_h$ and $m_{H_0}$, the charged scalar mass is $m_{H^\pm}$ and the CP-odd scalar mass is $m_A$. At the superpartners mass scale $M_{SUSY}$ the $m_A$ and $\tan \beta$ can be chosen as the input parameters which fix the dimension-two parameters $\mu^2_1,\mu^2_2$ and $\mu^2_{12}$ of the Higgs potential, while  the dimensionless factors $\lambda_i$ ($i=$1,...,7) at the tree level are expressed, using the $SU(2)$ and $U(1)$ gauge couplings $g_2$ and $g_1$, as follows~\cite{full_mssm_2, flores83}
\begin{eqnarray}
\label{lts}
\lambda_{1,2}^{\tt tree}(M_{SUSY})&=&\frac{g_1^2+g_2^2}{8}, \qquad
\lambda_{3}^{\tt tree}(M_{SUSY})=\frac{g_2^2-g_1^2}{4},\nonumber \\
\lambda_4^{\tt tree}(M_{SUSY})&=&-\frac{g_2^2}{2}, \qquad
\lambda_{5,6,7}^{\tt tree}(M_{SUSY})=0.
\label{boundary_condition}
\end{eqnarray}
The dimension-two parameters $\mu^2_1,\mu^2_2$ and $\mu^2_{12}$ are fixed using the minimization conditions:
\begin{eqnarray}
\mu_1^2 &=&{} -m_A^2 \sin^2(\beta)+\frac{m_Z^2}{2} \cos(2 \beta), \nonumber \\
\mu_2^2 &=&{} -m_A^2 \cos^2(\beta)-\frac{m_Z^2}{2} \cos(2 \beta), \nonumber  \\
\mu_{12}^2 &=& m_A^2 \sin(\beta) \cos(\beta), \label{mu}
\end{eqnarray}
where $m_Z=v \, \sqrt{g_1^2+g_2^2}/2$. Then the potential given by eq.~(\ref{genU})
can be rewritten in the mass basis of scalar bosons, which are massless Goldstone bosons $G^0$, $G^+$, $G^-$ and massive Higgs bosons $h$, $H_0$, $A$, $H^+$, $H^-$:
\begin{equation}
\label{Uphys}
V(h,H_0,A,H^\pm,G^0,G^\pm)=\frac{m_h^2}{2} h^2+\frac{m^2_{H_0}}{2} H_0^2+\frac{m_A^2}{2} A^2+m_{H^\pm}^2 H^+ H^- +I_3+I_4,
\end{equation}
where
\begin{eqnarray}
m_h^2 &=& m_Z^2 \sin^2 (\alpha+\beta)+m_A^2 \cos^2(\alpha-\beta), \\
m_{H_0}^2 &=& m_Z^2 \cos^2 (\alpha+\beta) +m_A^2 \sin^2 (\alpha-\beta), \\
m_{H^\pm}^2 &=& m_A^2+m_W^2.
\end{eqnarray}
Explicit forms of the interaction terms $I_3$ and $I_4$ are presented in the appendix~\ref{apB}.
The mixing angles $\alpha$ and $\beta$ at the $M_{SUSY}$ scale are connected by the following equation
\begin{equation}
\label{alpha}
\tan(2\alpha)=\frac{m_A^2+m_Z^2}{m_A^2-m_Z^2} \tan (2 \beta).
\end{equation}
The scalar resonance with mass $125\, GeV$ which is experimentally observed at the LHC~\cite{experiment_0, experiment_1, experiment_2} has properties consistent with the SM. However, MSSM identifications are still possible with limited experimental statistics. Experimental data of the LHC Run I demonstrates the SM-like couplings of observed Higgs boson to fermions and vector bosons at the level of statistical significance only on the level slightly better than 2$\sigma$~\cite{agreement}. In the following consideration, the CP-even state $h$ of the MSSM, when it is overridden mass, which is determined by the radiation corrections from the squark sector, will be identified as the $125 \,GeV$ resonance. In the presence of other scalars $H_0$, $A$, $H^+$ and $H^-$, which are not experimentally observed, such identification is possible for the two specific features in the MSSM parameter space: (i) the decoupling regime~\cite{decoupling} and/or (ii) the alignment limit~\cite{alignment_0, alignment_1}.
In the decoupling regime masses of scalars $H_0$, $A$, and $H^\pm$ are very large (they are at multi-TeV scale where also the lightest superpartners can be found), so their contributions to the observables at the top quark scale are strongly suppressed, while in the alignment limit  $H_0$, $A$, and $H^\pm$ are not necessarily extremely heavy. The alignment limit will be used in the following consideration.
In this limit $\beta-\alpha\approx \pi/2$  and the potential in eq.~(\ref{genU}) can be simplified by a special choice of mixing angles $\alpha$ and $\beta$. After rotation of scalar isodoublets
\begin{equation}
\Phi^{'}_1={}-\Phi_1 \sin \beta+ \Phi_2 \cos \beta, \quad \Phi^{'}_2=\Phi_1 \cos \beta+ \Phi_2 \sin \beta
\end{equation}
to so-called Higgs basis~\cite{full_mssm_2}
and the choice of mixing angles $\beta=\pi/2$ and $\alpha=0$, the $SU(2)$ components of isodoublets and the vacuum expectation values are
\begin{equation}\label{Simplific}
  \eta_1=H_0, \quad \eta_2=h, \quad  v_1=0,\quad v_2=v.
\end{equation}
So, in the unitary gauge $G^0=G^\pm=0$, we get
\begin{equation}\label{newvars}
    \chi_1={}-A, \quad \chi_2=0, \quad  \omega_1^\pm={}-H^\pm, \quad \omega_2^\pm=0
\end{equation}
and the isodoublet convolutions are given by
\begin{equation}
\label{psihu}
        (\Phi_1^\dag\Phi_1) =H^-H^+ +\frac{A^2}{2}+\frac{H^2_0}{2}\equiv\frac{1}{2}(\Omega^2_\pm+\Omega^2_0), \qquad
       (\Phi_2^\dag\Phi_2)=\frac{h_v^2}{2},
\end{equation}
\begin{equation}
(\Phi_1^\dag\Phi_2)=\frac{h_v}{2}(H_0+iA),\qquad (\Phi_2^\dag\Phi_1)=\frac{h_v}{2}(H_0-iA),
\end{equation}
where $h_v=h+v$, $\Omega^2_0={H_0}^2+A^2$, and $\Omega^2_\pm=2 \, H^+ H^-$. The kinetic terms have canonical form
 \begin{equation}
   \partial_\mu \Phi_1^\dag\partial^\mu \Phi_1 =  \partial_\mu H^- \partial^\mu H^{+} +\frac{1}{2} \, (\partial A)^2 +\frac{1}{2}(\partial H_0)^2, \quad  \partial_\mu \Phi_2^\dag\partial^\mu \Phi_2=\frac{1}{2} \, (\partial h)^2.
        \end{equation}

It follows from eq.~(\ref{mu}) that $\mu^2_{12}=0$ and the potential in eq.~(\ref{Uphys})
becomes
\begin{eqnarray}
V(h_v,\Omega_0,\Omega_\pm)&=&{}- m^2_1 h^2_v + m^2_2 \left( \Omega^2_0 +  \Omega^2_\pm \right) \nonumber \\
&+&\nu_1 ( h^4_v +  \Omega^4_0 + \Omega^4_\pm )
- 2\nu_{1} h^2_v \Omega^2_0 + 2\nu_{2} h^2_v \Omega^2_\pm + 2\nu_{1} \Omega^2_0 \Omega^2_\pm\vspace{0mm}, \label{inspired}
\end{eqnarray}
where
\begin{equation*}
m^2_1=\frac{m^2_Z}{4}, \quad m^2_2=\frac{m^2_A}{2}+\frac{m^2_Z}{4},
\end{equation*}
\begin{equation*}
\nu_1=\frac{g^2_1+g^2_2}{32}, \quad \nu_2=\frac{g^2_2-g^2_1}{32}.
\end{equation*}

The potential in eq.~(\ref{inspired}) qualitatively corresponds to the MSSM potential at the scale $M_{SUSY}$. It is invariant under two-dimensional rotations in ($H_0$, $A$) space and ($H^+$, $H^-$) space, what is the consequence of the specific choice of the mixing angles $\alpha$ and $\beta$ in the alignment limit. This property allows reducing the number of five physically significant fields $h$, $H_0$, $A$, $H^+$ and $H^-$ to the three field combinations, $h^2_v$, $\Omega^2_0$ and $\Omega^2_\pm$. Note that at $h_v=0$ the potential given by eq.~(\ref{inspired}) is invariant under rotations in the four-dimensional field space.
Tree-level quartic couplings $\lambda_i$, eq.~(\ref{boundary_condition}), are expressed through the gauge couplings $g_{1,2}$ which are fixed by collider data, since the gauge boson masses at tree level $m_Z=v\sqrt{g^2_1+g^2_2}/2$, $m_W=v\, g_2/2$ and cross sections of $W^\pm$, $Z$ production are precisely measured ($v=\sqrt{v^2_1+v^2_2}=(G_F\sqrt{2})^{-1/2}$, $G_F$ is the Fermi constant). Substituting $m_Z=91.2 \,GeV$ and $m_W=80.4 \,GeV$, we obtain $g_1=0.36$ and $g_2=0.65$, which are used in numerical calculations of section~\ref{sec:5}.


\section{The MSSM-inspired model with non-minimal interaction}\label{sec:3}

Generic action which is dependent on $N$ scalar fields $\phi^I$, $I=1, ..., N$ with the standard kinetic term and non-minimal coupling to gravity can be written as
\begin{equation}
S_{J}=\int d^4x\sqrt{-\tilde{g}}\left[f(\phi^I)\tilde{R}-\frac12\delta_{IJ}\tilde{g}^{\mu\nu}\partial_{\mu}\phi^I\partial_{\nu}\phi^J- V(\phi^I) \right],
\label{Jordan action}
\end{equation}
where tilde denominates the metric tensor and curvature in the Jordan frame. In our case $V(\phi^I)$ depends on five real scalar fields
\begin{equation}
\phi^1=\frac{H^+ +H^-}{\sqrt{2}},\, \phi^2=\frac{H^+ -H^-}{\sqrt{2}\mathrm{i}},\, \phi^3=A, \, \phi^4=H_0,\, \phi^5=h_v.
\end{equation}
This action can be transformed to the following action in the Einstein frame~\cite{Kaiser:2010ps} (see also~\cite{KaiserHiggs,Kaiser}):
\begin{equation}
\label{SE}
S_{E}=\int d^4x\sqrt{-g}\left[\frac{M^2_{Pl}}{2}R-\frac12\mathcal{G}_{IJ}{g^{\mu\nu}}\partial_\mu\phi^I\partial_\nu\phi^J-W\right],
\end{equation}
where
\begin{equation*}
\mathcal{G}_{IJ}=\frac{M^2_{Pl}}{2f(\phi^K)}\left[\delta_{IJ}
+\frac{3f_{, I} f_{, J}}{f(\phi^K)}\right],\qquad W= M^4_{Pl}\frac{V}{4f^2},
\end{equation*}
the reduced Planck mass $M_{Pl}\equiv1/\displaystyle\sqrt{8\pi G}$, $f_{,I} = \partial f/\partial \phi^I$. Metric tensors
in the Jordan and the Einstein frames are related by the equation
\begin{equation*}
g_{\mu\nu}=\frac{2}{M^2_{Pl}}f(\phi^I)\tilde{g}_{\mu\nu}.
\end{equation*}
In the single-field Higgs-driven inflation the function $f$ has been chosen as a sum of the Hilbert--Einstein term and the induced gravity term. We choose the function $f$ in an analogous form:
\begin{equation}
\label{fp2}
    f(\Phi_1,\Phi_2)=\frac{M^2_{Pl}}{2}+\xi_1\Phi_1^\dagger\Phi_1+\xi_2\Phi_2^\dagger\Phi_2,
\end{equation}
where $\xi_1$ and $\xi_2$ are positive dimensionless constants.
This form of function $f$ follows from the requirement of renormalizability for quantum field theories in curved space-time~\cite{Chernikov_0, Chernikov_1,Callan:1970ze,BOS}, where non-minimal couplings appear as renormalization counterterms for scalar fields. We also assume that vacuum expectation values for scalar fields are negligibly small in comparison with $M_{Pl}$.

Note that non-minimal interaction in the form of eq.~(\ref{fp2}) was considered \cite{gong} in the framework of the (nonsuperymmetric) two-Higgs-doublet model, when the boundary condition eq.~(\ref{boundary_condition}) is not used and the Higgs potential includes seven quartic couplings. Arbitrariness of the choice of $\lambda_i$ is constrained imposing exact or approximate $Z_2$ symmetry (discrete symmetry whose breaking results in the appearance of the axion) on the generic two-Higgs-doublet potential which takes a specific functional form different from the 'MSSM-inspired', eq.~(\ref{inspired}). It is assumed that the Higgs doublets $\Phi_1$ and $\Phi_2$ in this simplified potential are $(0,v_1/\sqrt{2}$) and $(0,v_2/\sqrt{2})$, what happens if the fields $\omega_{1,2}$, $\eta_{1,2}$ and $\chi_{1,2}$ in eq.~(\ref{dublets}) are taken to be zero, so the MSSM mass eigenstates $h$, $H$, $A$ and $H^\pm$ are not specified. In our case the function $f$ depends on the five scalar fields:
\begin{equation}
f(\Phi_1,\Phi_2)=\frac{M^2_{Pl}}{2}+\frac{\xi_1}{2}(\Omega^2_{\pm}+\Omega^2_0)+\frac{\xi_2}{2}h^2_v.
\end{equation}


\section{Properties of the equations of motion in the FLRW metric}\label{sec:4}

Let us consider a spatially flat FLRW universe with metric
interval
\begin{equation*}
ds^2={}-dt^2+a^2(t)\left(dx_1^2+dx_2^2+dx_3^2\right),
\end{equation*}
where $a(t)$ is the scale factor. Varying the action in eq.~(\ref{SE}) with respect to $g_{\mu \nu}$ and fields we get the following equations for the FLRW metric
\begin{equation}\label{Equations}
H^2=\frac{1}{3M_{Pl}^2}\left(\frac{\dot{\sigma}^2}{2}+W\right), \qquad
  \dot{H}={} -\frac{1}{2M_{Pl}^2}\dot{\sigma}^2,
\end{equation}
where the Hubble parameter $H=\dot{a}/a$, $\dot{\sigma}^2=\mathcal{G}_{IJ} \dot \phi^I \dot \phi^J$, and dots mean the time derivatives.
Field equations have the following form~\cite{Kaiser}
\begin{equation}
\label{EqFields}
\ddot\phi^I+3H\dot\phi^I + \Gamma^I_{\>\> JK}  \dot \phi^J \dot \phi^K + \mathcal{G}^{IK} W'_{, K} = 0\,,
\end{equation}
where $\Gamma^I_{\>\> JK}$ is the Christoffel symbol for the field-space manifold, calculated in terms of $\mathcal{G}_{IJ}$, $W{'}_{,K} = \partial W/\partial \phi^K$. Hereafter, primes denote derivatives with respect to the fields. Due to the relationship of inflationary evolution in the Jordan and the Einstein frames, eqs.~(\ref{Equations}) and (\ref{EqFields}) are equivalent to eqs.~(\ref{General_metric_equations}) and (\ref{General_fields_equations}) after transformation of the latter to Einstein frame.

During inflation the Hubble parameter is positive and the scalar factor is a monotonically increasing function. To describe the evolution of scalar fields during inflation we use the number of e-foldings $N_e=\ln(a/a_{e})$, where $a_e$ is the value of the scalar factor at the end of inflation, as a new measure of time.
The notation $N^*_e=-N_e$ will be also used for convenience.

Using ${d}/{dt}=H\, {d}/{dN_e}$ one can write eqs.~(\ref{Equations}) and (\ref{EqFields}) in the form
\begin{eqnarray}
  &&H^2=\frac{2W}{6M_{Pl}^2-(\sigma^\prime)^2},  \label{H2N}\\
  &&\frac{d\ln{H}}{dN_e}={} -\frac{1}{2M_{Pl}^2}\left(\sigma^\prime\right)^2,\label{dlnHN}\\
  &&\frac{d\phi^I}{dN_e}=\psi^I, \label{equpsi}\\
  &&\frac{d\psi^I}{dN_e}={} -\left(3+\frac{d\ln{H}}{dN_e}\right)\psi^I - \Gamma^I_{\>\> JK}\psi^J\psi^K -  	  \frac{1}{H^2}\mathcal{G}^{IK} W^{'}_{,K} , \label{diff eqN}
\end{eqnarray}
where $(\sigma^{'})^2 = H^2 \, (\dot{\sigma})^2$.
After substitution of eqs.~({\ref{H2N}) and (\ref{dlnHN}) a system defined by eqs.~(\ref{equpsi}) and (\ref{diff eqN}) includes ten first order equations which are suitable for numeric integration. Integration was performed by means of built-in subroutines of several computer algebra systems with cross-checks of results. Note that so far in this section and in the previous section \ref{sec:3} we have not made any approximations.

In order to calculate the observables, spectral index $n_\mathrm{s}$ and tensor-to-scalar ratio $r$, slow-roll parameters are introduced analogously to the single-field inflation
\begin{equation}
\epsilon = -\frac{\dot{H}}{H^2},\quad
\eta_{\sigma\sigma} = M_{Pl}^2 \frac{{\cal M}_{\sigma\sigma}}{W},
\end{equation}
where
\begin{equation}
{\cal M}_{\sigma\sigma} \equiv \hat{\sigma}^K \hat{\sigma}^J ({\cal D}_K {\cal D}_J W),
\end{equation}
$\sigma^I=\dot{\phi^I}/\dot{\sigma}$ is the unit vector in the field space and $\cal D$ denotes a covariant derivative with respect to the field-space metric, ${\cal D}_I\phi^J=\partial_I \phi^J+ \Gamma^J_{IK}\phi^K$. Then
the spectral index $n_\mathrm{s}$ and tensor-to-scalar ratio $r$ at the time when a characteristic scale (50--65 e-foldings before the end of inflation) is of the order of the Hubble radius  in the course of inflation, can be calculated using the single-field equations valid to lowest order in slow-roll parameters~\cite{Kaiser,c_perturbations}
\begin{equation}
n_\mathrm{s}=1-6 \epsilon+2 \eta_{\sigma \sigma}, \qquad
r=16 \epsilon.
\end{equation}


\section{Numerical solutions of the equations of motion}\label{sec:5}

The isosurfaces for the potential $W(h_v,\Omega_0,\Omega_\pm)$ (one of the three variables is fixed) are shown in figure~\ref{fig1}. At the fixed value of $h_v$, $\Omega_0$ or $\Omega_\pm$ of the order of $0.1$ (in Planck units) the saddle configuration of the surface is observed, as shown in figure~\ref{fig1}(a). A characteristic feature of $W$ which demonstrates ridges and gullies is shown in figure~\ref{fig1}(b). In the gullies evolution of the field system looks as an infinite expansion at the constant Hubble parameter. One can see that the slow-roll inflation is possible if the initial condition for $h_v$ or $\Omega_0$ is chosen in the vicinity of zero, which is equivalent to four nonzero values of the fields $(A,H_0,H^\pm)$ or three nonzero values of the fields $(h_v,H^\pm)$. Initial conditions for a number of successful inflationary scenarios of this sort are presented in table~\ref{tabl:sets}.
The evolution of fields superimposed on the Einstein-frame potential for the inflationary scenarios $A$ and $B$, see table~\ref{tabl:sets}, is shown in figure~\ref{3D}, where the dashed fragments of field trajectories correspond to the inflationary stage
when $0 \leq N_e^*\leq 65$.
If we assume that the number of e-foldings during inflation $N_e^*=65$, then
we get the initial conditions for inflationary trajectories presented in tables \ref{tabl_A} and \ref{tabl_B}.
For scenarios $A_1$, $A_2$, $A_3$, $B_3$, and $B_4$
(see table \ref{tabl:nsr}) the value of the Hubble parameter in the beginning of inflation is $H_{init}<3.6\cdot10^{-5} \, M_{Pl}$. Note that this value of the Hubble parameter is found to be in good agreement with the observational data~\cite{PlanckIfl_0, PlanckIfl_1}.

\begin{table}[htbp]
\begin{center}
\begin{tabular}{|c|cc|ccccc|}
\hline
Scenario & $\xi_1$ & $\xi_2$ & $\phi^1_0$ & $\phi^2_0$ & $\phi^3_0$ & $\phi^4_0$ & $\phi^5_0$ \\  \hline
$A_1$ & 2500 & any & 0.2 & 0.24 & 0.3 & 0.1 & 0
\\
$A_2$ & 2500 & any & 2$\times 10^{-3}$ & 0 & 0.45 & 0.2 & 0
\\
$A_3$ & 2500 & any & 0.2 & 0.26 & 0.5 & 0.6 & 0 \\
$A_4$ & 40 & any & 0.8 & 0.9 & 0.5 & 0.7 & 0
\\
$B_1$ & 1100 & 500 & 0.3 & 0.2 & 0 & 0 & 0.1
\\
$B_2$ & 1100 & 500 & 0.4 & 0.6 & 0 & 0 & 0.3
\\
$B_3$ & 2200 & 1000 & 0.3 & 0.2 & 0 & 0 & 0.1
\\
$B_4$ & 2200 & 2200 & 0.1 & 0.1 & 0 & 0 & 0.155
\\
\hline
\end{tabular}
\caption{ Initial conditions (in units of $M_{Pl}$) for trajectories with successful inflationary scenarios, CP-odd Higgs boson mass $m_A=200\, GeV$. The $SU(2)$ and $U(1)$ gauge couplings are $g_1=0.36$ and $g_2=0.65$.
}
\label{tabl:sets}
\end{center}
\end{table}
\vskip 0mm
\begin{table}[h]
\begin{center}
\begin{tabular}{|c|cccc|cccc|}
\hline
 Scenario & $\phi^1_{in}$ & $\phi^2_{in}$ & $\phi^3_{in}$ & $\phi^4_{in}$ & $\psi^1_{in}$  & $\psi^2_{in}$ & $\psi^3_{in}$ & $\psi^4_{in}$  \\ \hline
 $A_1$   & 0.0849 & 0.1019 & 0.1274 & 0.0425 & $-$0.0006 & $-$0.0008 & $-$0.0010 & $-$0.0003  \\
  $A_2$  & 0.0008 & 0 & 0.1725 & 0.0767 & $-$0.000006 & 0 & $-$0.0013 & $-$0.0006   \\
 $A_3$ & 0.0446 & 0.0579 &0.111 & 0.134 & $-$0.0003 & $-$0.0004 &  $-$0.0008 &  $-$0.0010  \\
 $A_4$ & 0.7984 & 0.8982 & 0.4990 & 0.6986 & $-$0.0049 & $-$0.0055 &  $-$0.0030 &  $-$0.0043  \\
 \hline
\end{tabular}
\caption{Initial conditions (fields in units of $M_{Pl}$) at $N_e^*=65$ in the scenarios of type $A$ (see table~\ref{tabl:sets}). }
\label{tabl_A}
\end{center}
\end{table}
\vskip 0mm
\begin{table}[h]
\begin{center}
\begin{tabular}{|c|ccc|ccc|}
\hline
 Scenario & $\phi^1_{in}$ & $\phi^2_{in}$ &$\phi^5_{in}$ & $\psi^1_{in}$ & $\psi^2_{in}$ & $\psi^5_{in}$  \\ \hline
 $B_1$ & 0.2367 & 0.1578 & $ 1.2 \cdot10^{-8}$ &  $-0.0018$& $-0.0012$ & $-7.7\cdot10^{-7} $ \\
 $B_2$ &0.1578 & 0.2367& $-9.9\cdot10^{-21}$ & $-0.0012$& $-0.0018$ & $-1.6\cdot10^{-20}$\\
 $B_3$ &0.1675 & 0.1117& $3.2\cdot10^{-17}$ & $-0.0013$& $-0.00084$ & $5.2\cdot10^{-17}$\\
 $B_4$ &0.1009 & 0.1009& $0.1426$ & $-0.000756$& $-0.000756$ & $-0.00107$\\
\hline
\end{tabular}
\caption{Initial conditions (fields in units of $M_{Pl}$) at $N_e^*=65$ in the scenarios of type $B$ (see table~\ref{tabl:sets}). }
\label{tabl_B}
\end{center}
\end{table}
\begin{table}[h]
\begin{center}
\begin{tabular}{|c|c|c|c|}
\hline
Scenario & $H$
[$10^{-5}$] & $r$ & $n_\mathrm{s}$\\  \hline
$A_1$ & 2.99983 &0.00266259 & 0.969398 \\
$A_2$ & 2.99983 & 0.00266259 & 0.969398 \\
$A_3$ & 2.99983 & 0.00266255 & 0.969399 \\
$A_4$ & 187.444 & 0.00174899 & 0.969258 \\
$B_1$ & 6.81778 & 0.00266322 & 0.969396\\
$B_2$ & 6.81778 & 0.00266325 & 0.969396 \\
$B_3$ &3.40892& 0.00265832 & 0.969424 \\
$B_4$ &2.98224& 0.00263611 & 0.969555 \\
\hline
\end{tabular}
\caption{The Hubble parameter $H$ (in units of $M_{Pl}$), tensor-to-scalar ratio $r$ and spectral index $n_\mathrm{s}$ for successful inflationary scenarios at $N_e^*=65$, $m_A=200 \,GeV$.}
\label{tabl:nsr}
\end{center}
\end{table}

\begin{figure}[h!tbp]
\centering
\begin{minipage}[h]{0.45\linewidth}
\center{\includegraphics[width=1\linewidth]{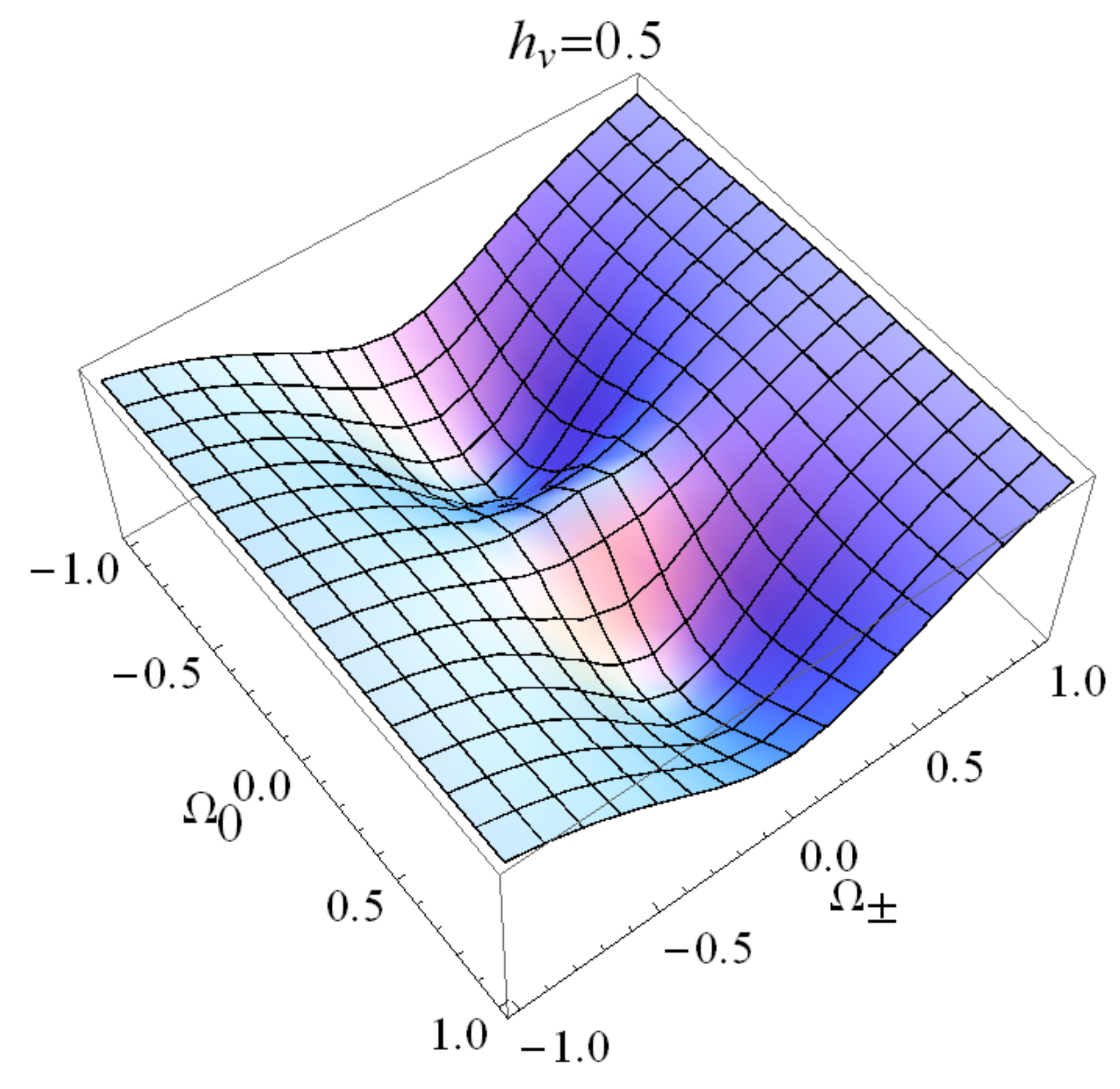}}\\(a)
\end{minipage}
\hfill
\begin{minipage}[h]{0.45\linewidth}
\center{\includegraphics[width=1\linewidth]{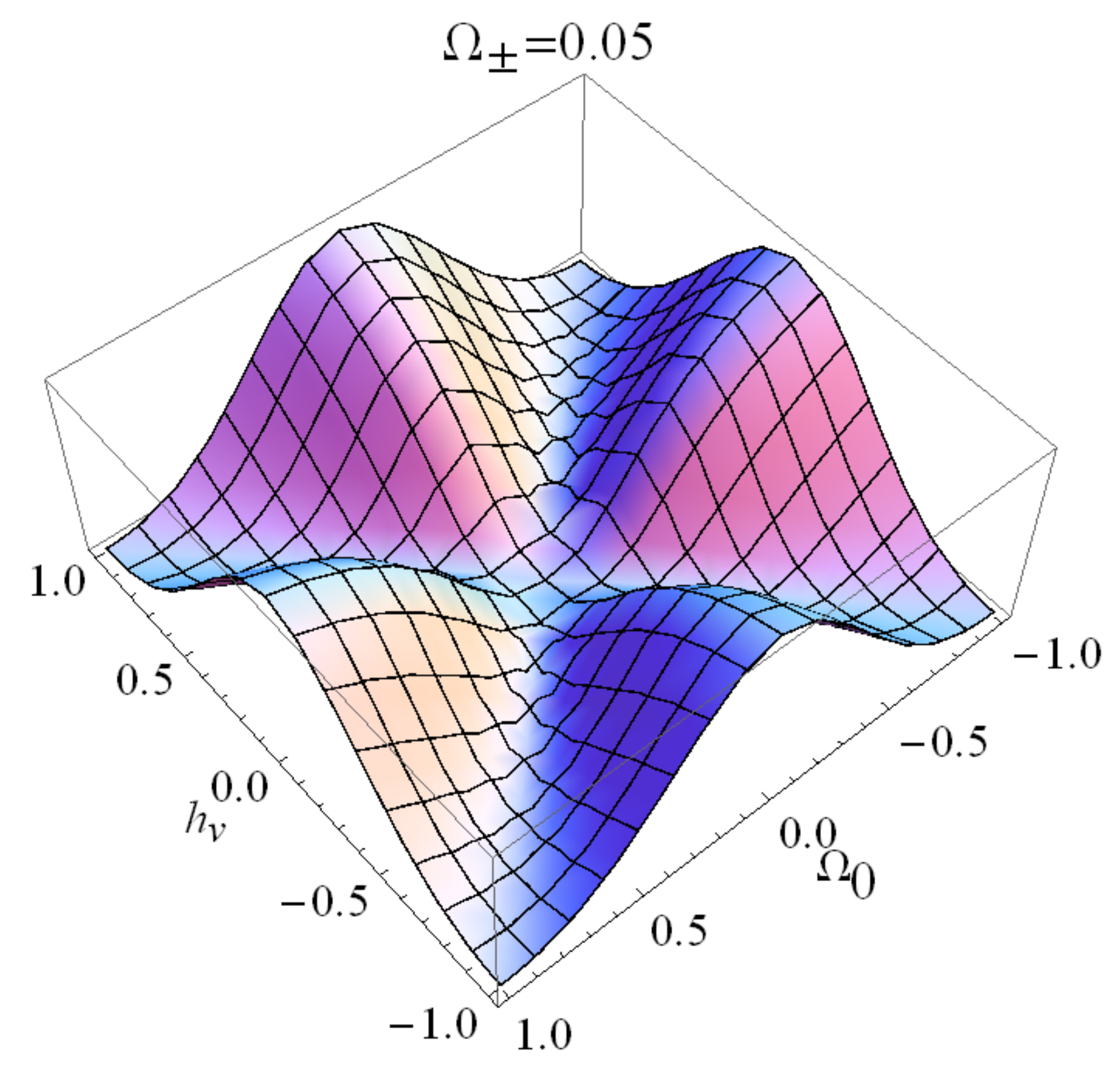}}\\(b)
\end{minipage}
\caption{\label{fig1} The isosurfaces of the potential $W(h_v,\Omega^2_0,\Omega^2_\pm)$ at fixed values of the field configurations (in units of $M_{Pl}$),
where $m_A=200 \,GeV$, $\xi_1=\xi_2=10$.}
\end{figure}

One can see that for the type $A$ inflationary scenarios the field system rolls slowly down to the potential minimum (see also figures~\ref{fields}(a) and \ref{fields}(b)), while for the $B$ type inflationary scenarios, except $B_4$, all nonzero fields demonstrate rapidly damped oscillations going to zero $h_v$
for the number of e-foldings before the end of inflation $N_e^* \gg 65$, see figures \ref{fields}(c) and \ref{fields}(d). At the same time, significant nonzero value of $h_v$ in the initial field configuration is suitable for inflation in the case $B_4$ when $\xi_1=\xi_2$. Note that inflationary scenarios with initial conditions denoted by $A$ and $B$ in tables \ref{tabl_A}, \ref{tabl_B}
demonstrate remarkable stability of slow-roll parameters $\epsilon$, $\eta_{\sigma \sigma}$ and observables $r$ and $n_\mathrm{s}$.
In different cases with the Hubble parameter $H \sim 10^{-5} M_{Pl}$ the values of $n_\mathrm{s}$ and $r$  coincide up to three digits.
Such "attractor behavior" when over a wide range of initial conditions the system evolves along the same trajectory in the course of inflation is known for single-field models~\cite{liddle}, but it is not an obvious observation, generally speaking, for multifield models. In this sense the phenomenological stability inherent to the single-field Higgs inflation is preserved for the multifield MSSM-inspired model under consideration.

The problem of perturbative unitarity violation at a large values of $\xi$ parameters mentioned in the Introduction may persist in the MSSM although an order of magnitude smaller values of $\xi$ appear in comparison with the SM Higgs inflation (except $A_4$ scenario, see Table \ref{tabl:sets}). While in the SM for the Higgs inflation a simple unitarity bound can be derived $E<M_{Pl}/\xi$ on the general basis of power-counting formalism for effective theory (for example, \cite{burgess}), in the MSSM-inspired models with several fields such a simple criteria is not reliable and the situation with partial wave unitarity is much more difficult. Recent analysis~\cite{kanemura} for the case of a general two-Higgs-doublet model without any discrete symmetry imposed on the scalar potential leads to non-trivial constraints on the masses and mixings which may depend on the scenario of new physics at a high energy scale.

\begin{figure}[h!tbp]
\begin{minipage}[h]{0.45\linewidth}
\center{\includegraphics[width=1\linewidth]{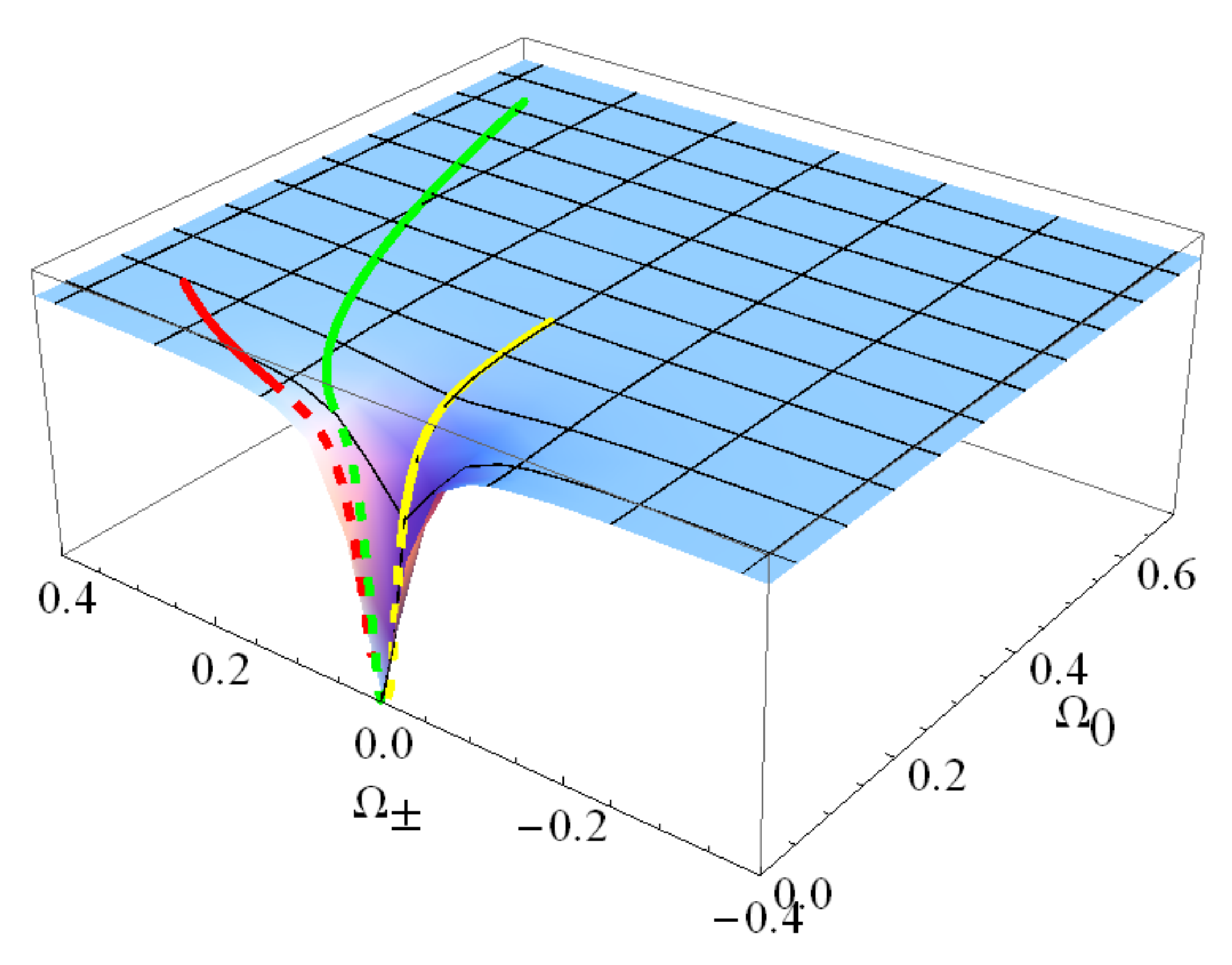}} \\(a)
\end{minipage}
\hfill
\begin{minipage}[h]{0.45\linewidth}
\center{\includegraphics[width=1\linewidth]{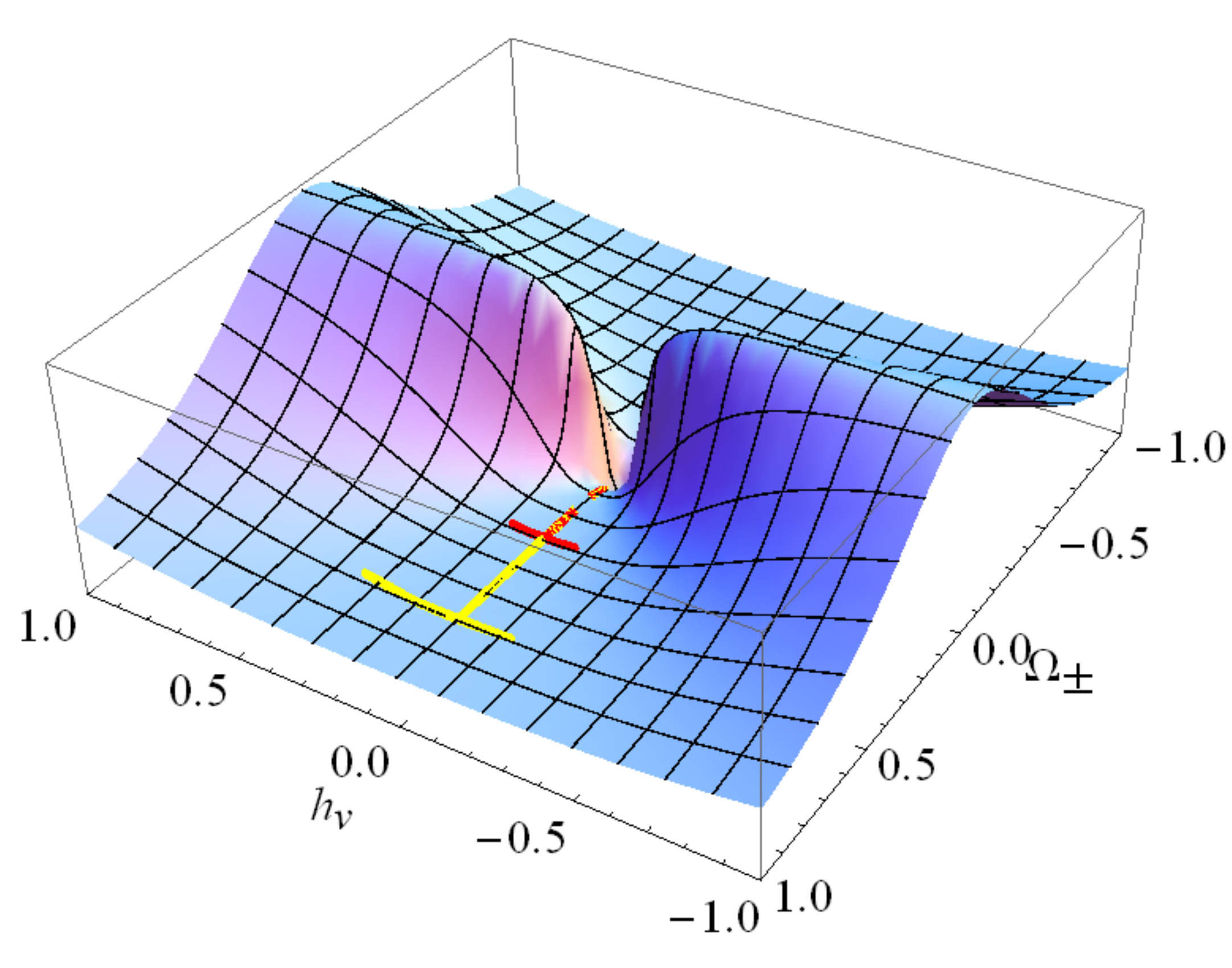}} \\(b)
\end{minipage}
\caption{\label{3D}
Parametric plots of the fields' evolution superimposed on the Einstein-frame potential for $A$ parameter sets, plot (a), and $B$ parameter sets, plot (b), see the parameter sets in table \ref{tabl:sets}. The trajectories shown here have the initial condition (in units of $M_{Pl}$):
(a) $\phi^5_0=0$ and $\phi^1_0=0.2$, $\phi^2_0=0.24$, $\phi^3_0=0.3$, $\phi^4_0=0.1$ (red line);
$\phi^1_0=2 \times 10^{-3}$, $\phi^2_0=0$, $\phi^3_0=0.45$, $\phi^4_0=0.2$ (yellow line);
$\phi^1_0=0.2$, $\phi^2_0=0.26$, $\phi^3_0=0.5$, $\phi^4_0=0.6$ (green line);
(b) $\phi^3_0=\phi^4_0=0$ and $\phi^1_0=0.3$, $\phi^2_0=0.2$, $\phi^5_0=0.1$ (red line);
$\phi^1_0=0.4$, $\phi^2_0=0.6$, $\phi^5_0=0.3$ (yellow line).
The dashed lines correspond to the inflationary stage when $0 \leqslant N_e^*\leqslant 65$.
}
\end{figure}
\begin{figure}[h!tbp]
\begin{minipage}[h]{0.48\linewidth}
\center{\includegraphics[width=1\linewidth]{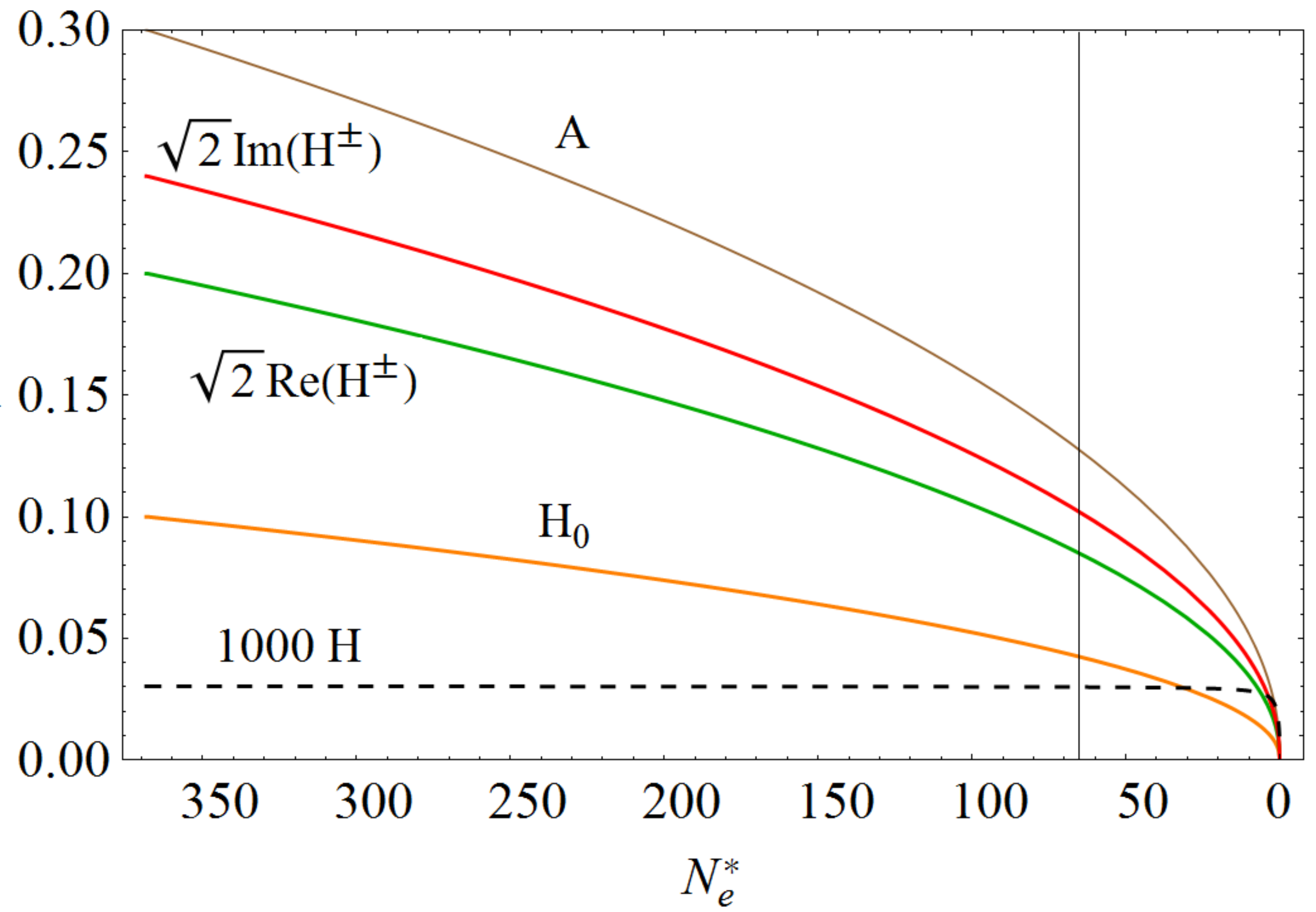}} \\(a)
\end{minipage}
\hfill
\begin{minipage}[h]{0.48\linewidth}
\center{\includegraphics[width=1\linewidth]{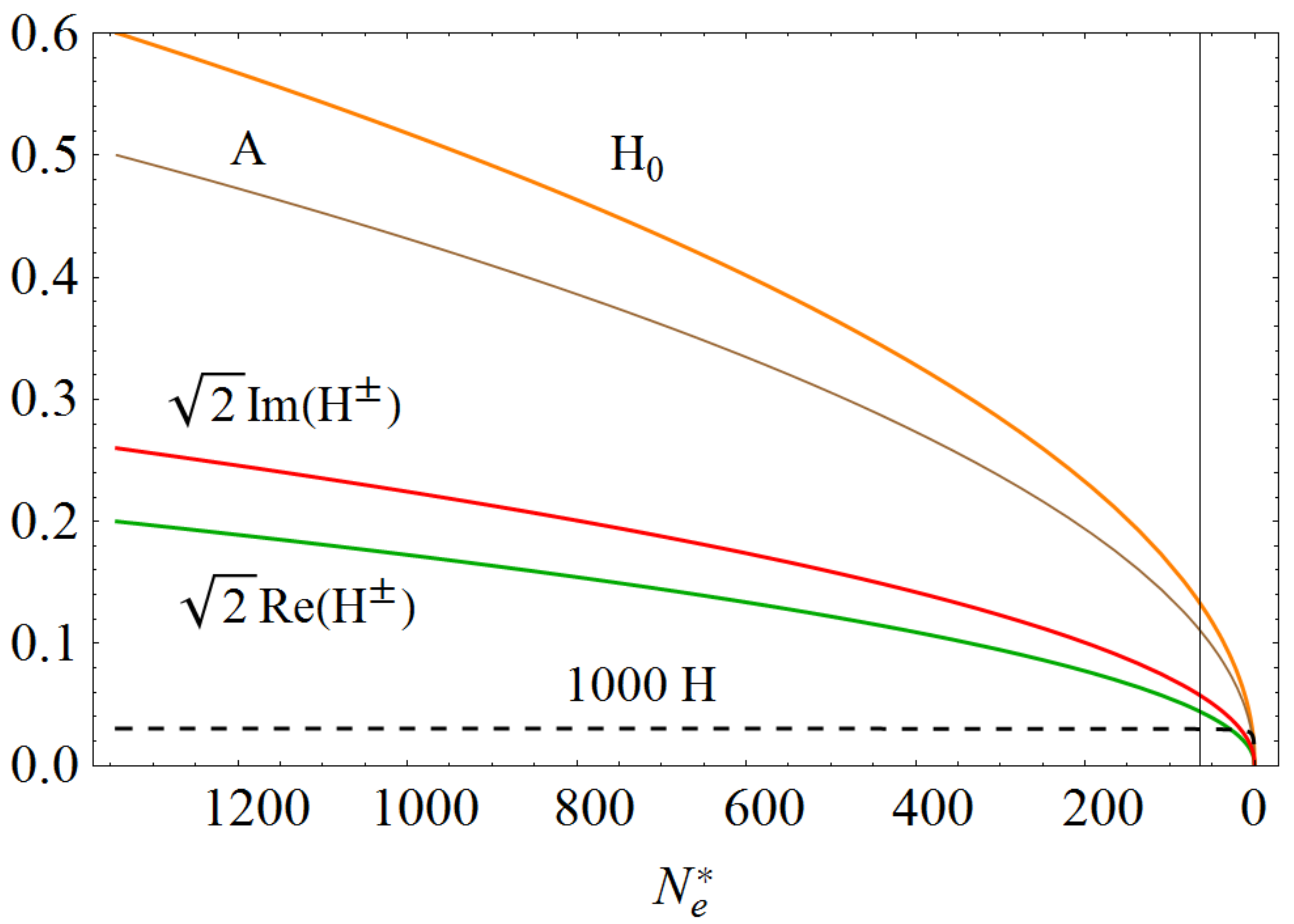}} \\(b)
\end{minipage}
\hfill
\begin{minipage}[h]{0.48\linewidth}
\center{\includegraphics[width=1\linewidth]{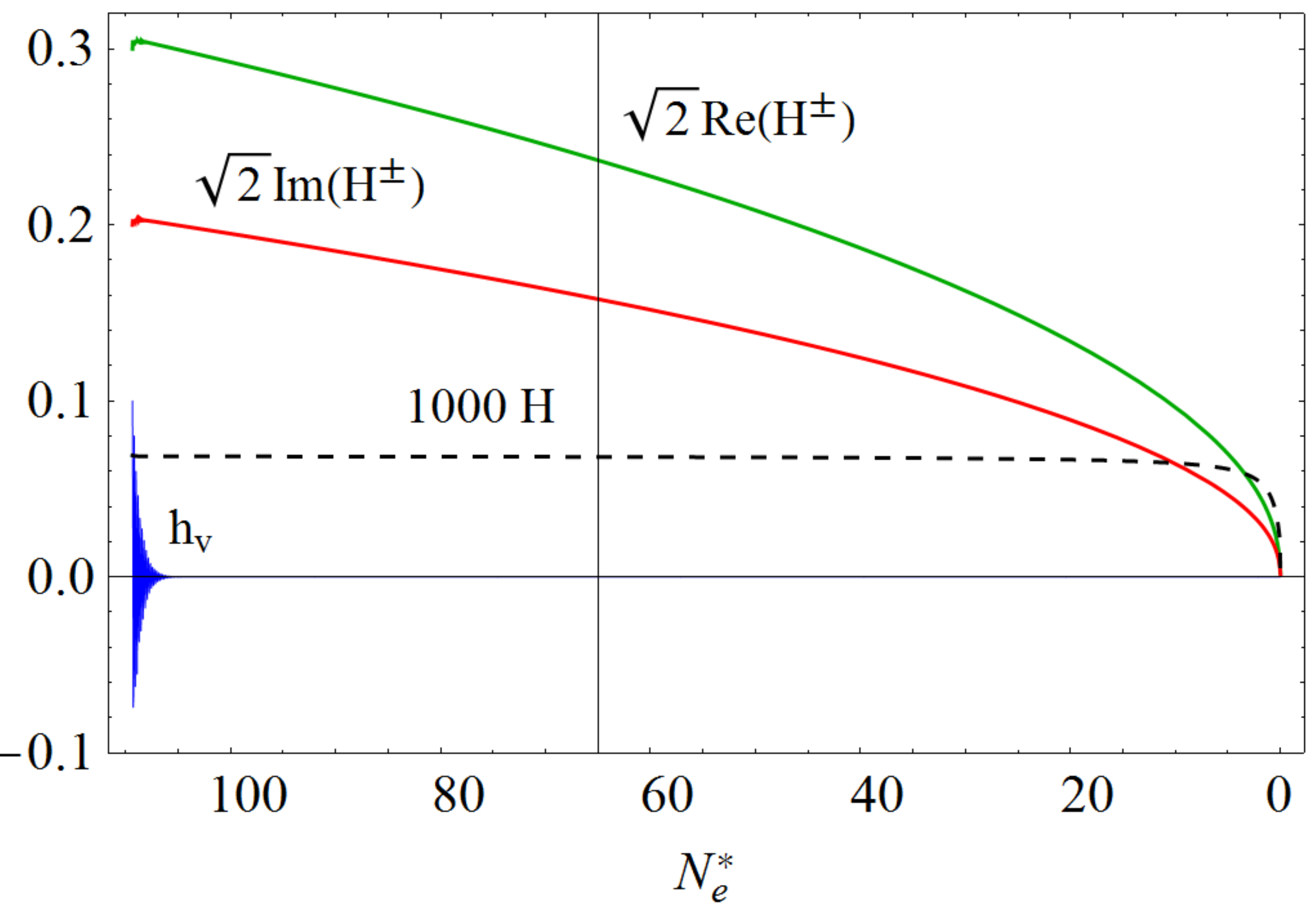}} \\(c)
\end{minipage}
\hfill
\begin{minipage}[h]{0.485\linewidth}
\center{\includegraphics[width=1.055\linewidth]{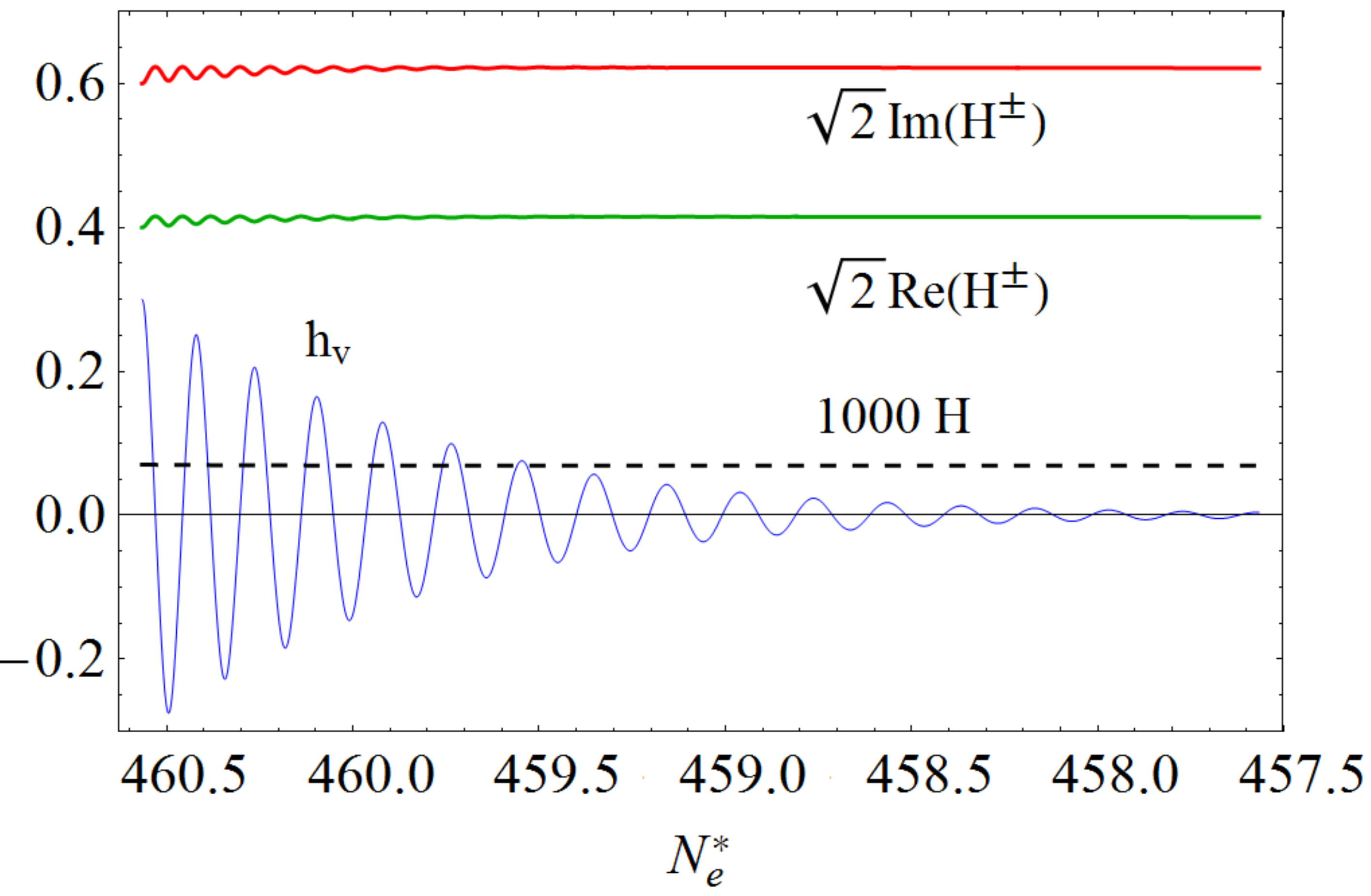}} \\(d)
\end{minipage}
\caption{\label{fields} Evolution of the fields and the Hubble parameter (in units of $M_{Pl}$) as functions of the number of e-foldings  before the end of inflation $N_e^*$ for the scenarios $A_1$ (a), $A_3$ (b), $B_1$ (c) and $B_2$ (d), see table \ref{tabl:sets}. The thin vertical black line corresponds to the number of e-foldings $N_e^*=65$. Plot (d) for the $B_2$ scenario shows the evolution for smaller scale in the $N^*_e$ interval from 460.5 to 457.5. The evolution (d) for the entire $N^*_e$ interval is similar to plot (c). }
\end{figure}
\begin{figure}[h!tbp]
\begin{minipage}[h]{0.5\linewidth}
\center{\includegraphics[width=0.9\linewidth]{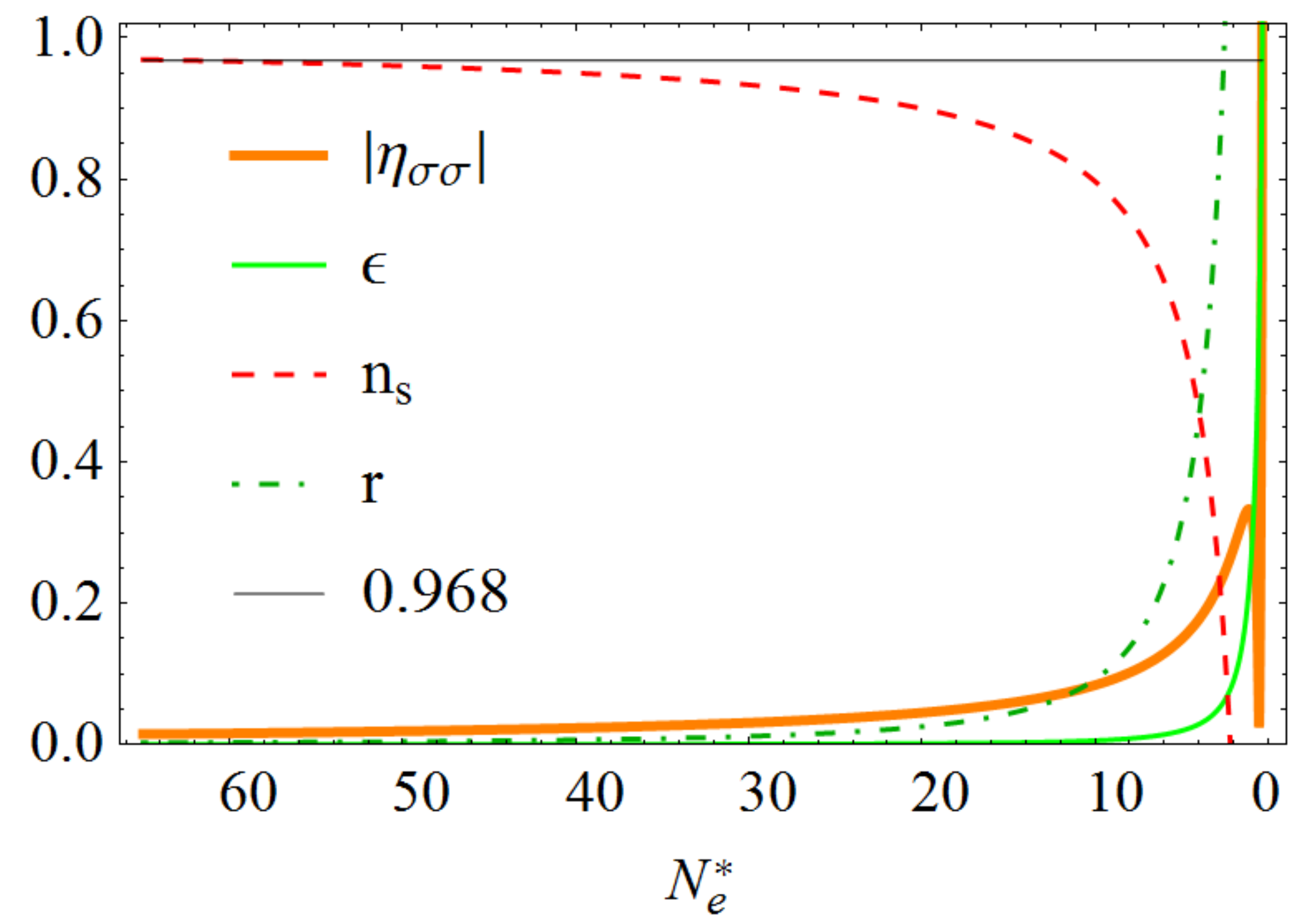}} \\(a)
\end{minipage}
\hfill
\begin{minipage}[h]{0.5\linewidth}
\center{\includegraphics[width=1\linewidth]{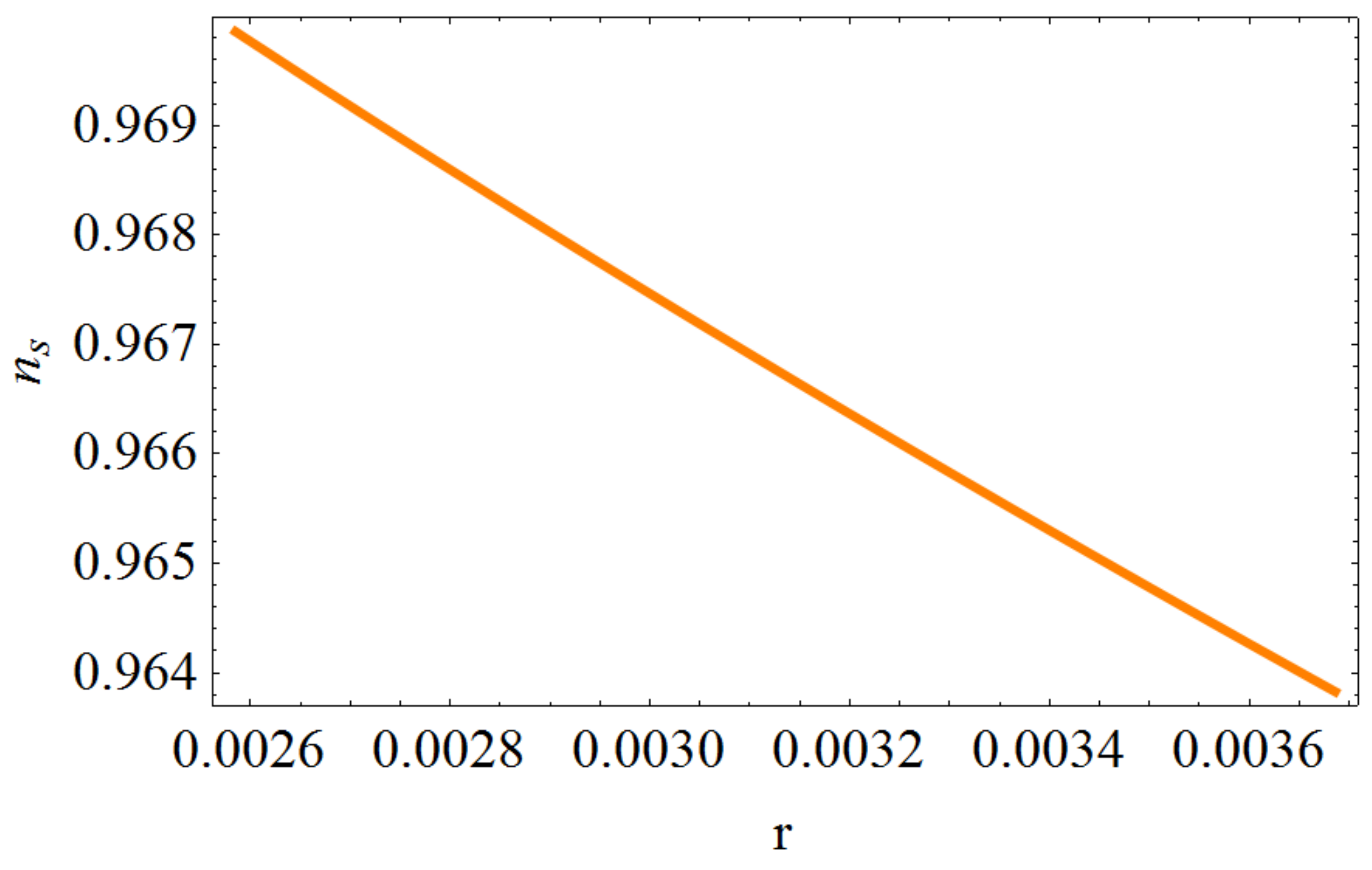}} \\(b)
\end{minipage}
\caption{\label{infl_par} (a) Evolution of the slow-roll parameters $\epsilon$ (light green line), $|\eta_{\sigma \sigma}|$ (dashed orange line), scalar-to-vector ratio $r$ (green line) and spectral index $n_\mathrm{s}$ (red line) as a number of e-foldings before the end of inflation $N_e^*$; (b)  ($n_\mathrm{s}$, $r$) contour at $55 \leqslant N_e^* \leqslant 65$.
}
\end{figure}

\section{The strong coupling approximation}\label{sec:6}

It follows from configurations shown in tables~\ref{tabl_A} and \ref{tabl_B} that quite different initial values of scalar fields and parameters $\xi_1$ and $\xi_2$ appear in all cases in combination with a very small value of $h_v$.
In all cases, but $A_4$, the inflationary parameters practically coincide, see figure~{\ref{infl_par}} and table \ref{tabl:nsr}.
In this section, we show that such a pattern can be explained in the framework of the so-called "strong coupling approximation". It has been shown in a large number of analyses \cite{Mukh, Roest:2013fha, LindeKallosh_etal_0, LindeKallosh_etal_1, LindeKallosh_etal_2, Binetruy:2014zya_0, Binetruy:2014zya_1, Ventury2015, EOPV2016} that there are several classes of the single-field inflationary models such that within a given class all models predict the same values of observable parameters $n_\mathrm{s}$ and $r$ in the leading $1/N_e$ approximation. These classes are known as \emph{cosmological attractors}. A similar analysis of two-field inflationary models has been made in~\cite{Kaiser:2013sna, Kallosh:2013daa}. For string-motivated supergravity theory in which both the field-space metric and the potential usually have poles at the same points, the inflationary dynamic and the corresponding attractor have been studied \cite{string_motivated}. The idea of a cosmological attractor is based on an observation that the kinetic term in Jordan frame practically does not affect the slow-roll parameters if the "strong coupling regime" is respected during inflation. In the case of multifield models  the field system is in the SC regime if the following inequality is respected:
\begin{equation}
\label{StrCC}
\delta_{IJ}\partial_\mu\phi^I\partial_\nu\phi^J\ll\frac{3}{f(\phi^K)}f_{,I}f_{,J}\partial_\mu\phi^I\partial_\nu\phi^J.
\end{equation}
In the approximation of eq.~(\ref{StrCC}) the action given in eq.~(\ref{SE}) can be written as
\begin{equation}
\label{SEapr}
S_{E}=\frac{M^2_{Pl}}{2}\int d^4x\sqrt{-g}\left[R-\frac{3g^{\mu\nu}}{2f^2(\phi^K)}f_{,I}f_{,J}\partial_\mu\phi^I\partial_\nu\phi^J
-\frac{M^2_{Pl}V(\phi^I)}{2f^2(\phi^I)}\right]
\end{equation}
and rewritten in the equivalent form
\begin{equation*}
S_{E}=\!\int\! d^4x\sqrt{-g}\left[\frac{M^2_{Pl}}{2}R-\frac{g^{\mu\nu}}{2}\partial_\mu\left[\sqrt{\frac32}M_{Pl}\ln\left(\frac{f}{f_0}\right)\right]
\partial_\nu\left[\sqrt{\frac32}M_{Pl}\ln\left(\frac{f}{f_0}\right)\right]
-\frac{M^4_{Pl}V}{4f^2}\right]\!,
\end{equation*}
where $f_0$ is a positive constant with the same dimension as $f$.
The role of inflaton in the strong coupling approximation is performed by the "effective field"
\begin{equation}
\label{lnf}
    \Theta=\sqrt{\frac32}M_{Pl}\ln\left(\frac{f}{f_0}\right),
\end{equation}
in terms of which the action $S_{E}$
includes the standard kinetic term of $\Theta$ and does not include kinetic terms of any other scalar fields which can be interpreted as model parameters. This circumstance allows one to calculate the inflationary parameters in the
SC approximation using the single-field model. If we adjust $\Theta$ in such a way that $\Theta=0$ corresponds to $\Omega_0=0$ and $\Omega_\pm=0$, then
$f_0={M^2_{Pl}}/{2}$.

The single-field model consistent with the above-mentioned scenarios $A$ and $B$ (see tables \ref{tabl_A} and~\ref{tabl_B}) can be easily defined.
In the scenario $A$ we set $\phi^5=h_v=0$ during inflation, while in the scenario $B$ (except $B_4$) one can observe that inflation starts when
\begin{equation*}
h_v^2\ll\sum\limits_{I=1}^4\left({\phi^I}\right)^2.
\end{equation*}
We do not consider the case $B_4$ here. In all other cases we can neglect $h_v$ and write the potential in the form
\begin{equation*}
    V_{sc}=m^2_2 \left(\Omega^2_0 +  \Omega^2_\pm \right)+\nu_1 \left(\Omega^2_0 + \Omega^2_\pm\right)^2.
\end{equation*}
The function $f$ is approximated by
\begin{equation}
    f_{sc}=\frac{M^2_{Pl}}{2}+\frac{\xi_1}{2}\left(\Omega^2_\pm+\Omega^2_0\right)
\end{equation}
and thereby
\begin{equation}
\label{Vappr}
    V_{sc}=\frac{m^2_2}{\xi_1} \left(2f-M^2_{Pl} \right)+\frac{\nu_1}{\xi_1^2} \left(2f-M^2_{Pl} \right)^2,
\end{equation}
so the Einstein frame  potential can be written as follows
\begin{equation}
\label{Wsc}
W_{sc}=\frac{M^4_{Pl}\left(M^2_{Pl}-2f_{sc}\right)[(M^2_{Pl}-2f_{sc})\nu_1-m_2^2\xi_1]}{4f_{sc}^2\xi_1^2}.
\end{equation}
Using $m_2^2\xi_1\ll M^2_{Pl}\nu_1$ we get from eq.~(\ref{Wsc})
\begin{equation}
W_{sc}\simeq\frac{M^4_{Pl}\nu_1}{\xi_1^2}\left(\frac{M^2_{Pl}}{2f_{sc}}-1\right)^2=\frac{M^4_{Pl}\nu_1}{\xi_1^2}
\left(1-\frac{M^2_{Pl}}{2f_0}e^{-\sqrt{6}\Theta/(3M_{Pl})}\right)^2.
\end{equation}
The slow-roll parameters are
\begin{equation*}
\epsilon = \frac{M^2_{Pl}}{2} \left(\frac{W^\prime_\Theta}{W}\right)^2=\frac{4}{3}\left(e^{\sqrt{6}\Theta/(3M_{Pl})}-1\right)^{-2},
  \quad    \eta=M^2_{Pl}\frac{W^{\prime\prime}_{\Theta}}{W}
  =\frac{4\left(e^{\sqrt{6}\Theta/(3M_{Pl})}-2\right)}{3\left(e^{\sqrt{6}\Theta/(3M_{Pl})}-1\right)^{2}}.
\end{equation*}
With these analytic expressions for the slow-roll parameters in the SC approximation the inflationary parameters can be easily calculated. It is convenient to express the inflationary parameters as a functions of $f_{sc}$
\begin{equation}\label{nsrf}
   n_\mathrm{s}=1-\frac{8M^2_{Pl}\left(M^2_{Pl}+2f_{sc}\right)}{3\left(M^2_{Pl}-2f_{sc}\right)^2},\qquad r=\frac{64M^4_{Pl}}{3\left(M^2_{Pl}-2f_{sc}\right)^2}\,.
\end{equation}
Straightforward numerical cross-checks demonstrate that the ratio
\begin{equation}
\label{Scc}
C_{sc}=\left|\frac{f(\phi^K)\delta_{IJ}\dot{\phi}^I\dot{\phi}^J}{3f_{,I}f_{,J}\dot{\phi}^I\dot{\phi}^J} \right |
\end{equation}
is less than $7\times 10^{-5}$ in the scenario $A$ and $2\times 10^{-4}$ in the scenario $B$, so the
SC approximation is meaningful. It is demonstrated in table~\ref{tabl_f} that in all cases the values of inflationary parameters $r$ and $n_\mathrm{s}$ calculated using eq.~(\ref{nsrf}) are close to the parameter values that have been found numerically in section \ref{sec:5}. Note in this connection that the primordial non-Gaussianities which do not arise in the single-field inflationary models should be very small in the case under consideration as soon as the reduction to a single-field scenario is precise enough.
It should be mentioned that $f_{in}/M^2_{Pl}$ close to 44 is not a sufficient condition for an inflationary scenario with suitable values of $n_s$ and $r$. For a large number of initial data with such values of $f_{in}/M^2_{Pl}$, but beyond the abovementioned $A$ and $B$ type scenarios, acceptable inflationary evolution is not observed\footnote{The initial conditions $\Omega_{\pm}=0$ and both $h_v$ and $\Omega_0$ nonzero lead to exotic situation when the field trajectory rapidly  (after $\sim 0.05$ e-foldings) rolls into the gully $h_v^2 = \Omega_0^2$ (see figure \ref{fig1}(b)). This direction is not absolutely flat (the case when critical points are degenerate and not isolated \cite{dp15}), but so close to flat that cannot be analysed by numerical methods. Simple estimate with $f_{in} = 45$ and $\xi_1+\xi_2 \sim 2\cdot 10^3$ gives an extremely long slow-roll with the number of e-foldings of the order of 10$^{12}$.}.

\begin{table}[h]
\begin{center}
\begin{tabular}{|c|c|c|c|}
\hline
Scenario & $f_{in}/M^2_{Pl}$ & $r$ & $n_\mathrm{s}$\\
\hline
$A_1$ & 43.346 & 0.00291& 0.96815 \\
$A_2$ & 44.834 & 0.00271 & 0.96925 \\
$A_3$ & 44.937& 0.00270 & 0.96932 \\
$A_4$ & 44.125 & 0.00280 & 0.96874 \\
$B_1$ & 45.123 & 0.00268 & 0.96945\\
$B_2$ & 45.024 & 0.00269 & 0.96938  \\
$B_3$ & 45.266 & 0.00266 & 0.96955 \\
\hline
\end{tabular}
\caption{The inflationary parameters in the strong coupling approximation calculated at $h_v=0$, $m_Z=0$ and $m_A=0$.}
\label{tabl_f}
\end{center}
\end{table}

\section{Summary}\label{sum}
In this paper, we constructed a MSSM-inspired extension of the original Higgs-driven inflation \cite{higgsinf_0, higgsinf_1, higgsinf_2, higgsinf_3, higssinflRG_0, higssinflRG_1, higssinflRG_2} using the two-Higgs-doublet potential of the MSSM which is simplified in a way suitable for calculation of transparent symbolic and numerical results for the main observables, the spectral index $n_\mathrm{s}$ and the tensor-to-scalar ratio $r$. The shape of the MSSM potential surface in the Einstein frame where ridges and bumps influence the trajectory in the fields space is different from the usual form in models of hybrid inflation. The model under consideration incorporates multiple non-minimally coupled scalar fields and non-canonical kinetic terms in the Einstein frame which are induced by the curvature of the field-space manifold. For these reasons, the evolution of fields is generically different from slow-roll, at least for some time interval during inflation.

The analysis of the background inflation dynamics demonstrated that after setting up the initial conditions for the five-dimensional field configuration such simplified MSSM-inspired model successfully describes the Higgs-driven inflation consistently with the observations of the Planck and BICEP2 collaborations. Two types of consistent inflationary scenarios are found with the initial conditions denoted as $A$ and $B$, see table \ref{tabl:sets}, which demonstrate the remarkable stability of the observables with respect to the shift of the initial field system configuration. The main difference between these two cases is the presence of rapid field oscillations in the initial phase of case $B$ before the beginning of inflation, while oscillations are absent in case $A$. During the period of cosmological evolution which determines the observables, $h_v$ field is negligibly small so the value of $\xi_2$ parameter practically does not influence the result and in the MSSM-inspired model degenerate values of $\xi_{1}$ and $\xi_{2}$ are always meaningful. Inflation occurs for field values much smaller than the Planck scale, although no suitable expansion scenario was found for initial state when $h_v$, $\Omega_0$ and $\Omega_\pm$ are very small at the same time. In all cases trajectories of the system do not turn steeply in the field space, so specific features of the potential like bumps and ridges are not expected to induce primordial non-Gaussianities with a magnitude large enough to be detectable in the cosmic microwave background.

Multifield model under consideration demonstrates rather strong attractor behavior and can be mapped to the single-field model with the effective inflaton field defined by eq.~(\ref{lnf}). Such models share very close results for the spectral index and the tensor-to-scalar ratio in combination with negligible non-Gaussianity, which are in good agreement with the latest experimental data.

In conclusion let us also note that an important point beyond our analysis is the stability of results with respect to radiative corrections. The flatness of the effective potential in the region of the field amplitudes of the order of $M_{Pl}$ is an essential property for a suitable slow-roll. While the quantum gravity corrections are expected to be rather small of the order of $V/M^4_{Pl} \sim g^2_p/ \xi^2$, the corrections induced by the SM fields and the superpartner fields involved in the $F$ and $D$ soft supersymmetry breaking Lagrangian terms require careful analysis which is dependent on the MSSM parametric scenario under consideration. For example, in the "natural MSSM scenario" which is used for LHC analyses the superpartners of quarks show up at the multi-TeV scale, while gauginos decouple. At the one-loop resummed level the superpartner threshold corrections to the two-doublet MSSM potential are expressed by Coleman-Weinberg terms $\Delta V = 1/(64\pi^2) Sp[(V^{''}(\phi))^2 (\log (V^{''}(\phi)/\mu^2)-3/2)]$, where second derivatives taken at the local minimum of Higgs potential are equal to masses of scalars. Nontrivial significant contributions are provided in the higher orders of perturbation theory by nonrenormalizable operators \cite{last}. Fermionic and bosonic loops give contributions of different signs which could partially compensate each other. Contributions of the SM vector bosons and fermions are smaller than the MSSM ones because of small gauge and Yukawa couplings, so main corrections from the third generation of quark superpartners interacting with Higgs isodoublets must not spoil a small slope of the potential. Important correction can be provided also by the renormalization group (RG) evolution of $\xi$ non-minimal couplings from the top quark scale to the $M_{Pl}$ scale. RG evolution gives at least a factor of two for the value of $\xi$ in the framework of SM Higgs-driven inflation, but moderate changes of the order of ten percent in the inflationary region. Models which are described by the RG-improved effective action~\cite{rgingl_0, rgingl_1, rgingl_2} should provide an improved precision for observables. Careful MSSM evaluations which are beyond our analysis are appropriate in order to ensure stability of results.

\acknowledgments
This work was partially supported by Grant No. NSh-7989.2016.2.
The research of E.O.P. and E.Yu.P. was supported in part by Grant No. MK-7835.2016.2.

\appendix
\section{General action for non-minimal Higgs interactions in the MSSM}
\label{apA}
In the general case one can write the action for non-minimal interaction of the MSSM Higgs doublets with gravity
in the form (here we redefine $f(\tilde{\Phi}_1,\tilde{\Phi}_2)=\frac{M^2_{Pl}}{2}[1+\varrho(\tilde{\Phi}_1,\tilde{\Phi}_2)]$)
  \begin{equation}
           S = \int{d^4x\sqrt{-g}\left\{ \frac{M^2_{Pl}}{2}\left[1+\varrho(\tilde{\Phi}_1,\tilde{\Phi}_2)\right]R - g^{\mu{}\nu{}}G^{IJ}\partial_{\mu}\tilde{\Phi}_I^\dag{}\partial_{\nu}\tilde{\Phi}_J-\mathcal{V}(\tilde{\Phi}_1,\tilde{\Phi}_2)\right\}}, \label{A1}
  \end{equation}
where $$ G^{IJ} = \begin{pmatrix} G^{11} & G^{12} \\ G^{21} & G^{22} \end{pmatrix},$$
     $$\varrho(\tilde{\Phi}_1,\tilde{\Phi}_2) = \sum_{a,b}{}\hat{\xi}_{ab}(\tilde{\Phi}_a^\dag\tilde{\Phi}_b) + \sum_{a,b,c,d}{}\hat{z}_{abcd}(\tilde{\Phi}_a^\dag\tilde{\Phi}_b)(\tilde{\Phi}_c^\dag\tilde{\Phi}_d)+...,$$
    $$\mathcal{V}(\Phi_1,\Phi_2) = -\sum_{a,b}{}\hat{\mu}_{ab}(\tilde{\Phi}_a^\dag\tilde{\Phi}_b) + \sum_{a,b,c,d}{}\hat{\lambda}_{abcd}(\tilde{\Phi}_a^\dag\tilde{\Phi}_b)(\tilde{\Phi}_c^\dag\tilde{\Phi}_d).$$
One can find some (may be non-unitary) transformation $\tilde{\Phi}_a \rightarrow\Phi_a = U_{ab}\tilde{\Phi}_b$ to diagonalize $G^{IJ}\rightarrow \delta^{IJ}$, so $U_{ab}$ is
\begin{equation*}
U_{ac}^\dag{}G^{cd}U_{db} = \delta_{ab}.
\end{equation*}

After such transformation the action can be written as
  \begin{equation}
           S = \int{d^4x\sqrt{-g}\left\{ \frac{M^2_{Pl}}{2}\left[1+\rho(\Phi_1,\Phi_2)\right]R - g^{\mu{}\nu{}}\delta^{IJ}\partial_{\mu}\Phi_I^\dag{}\partial_{\nu}\Phi_J-V(\Phi_1,\Phi_2)\right\}}, \label{A2}
  \end{equation}
where
\begin{equation*}
\rho(\Phi_1,\Phi_2) = \sum_{a,b}{}\xi_{ab}(\Phi_a^\dag\Phi_b)  + \sum_{a,b,c,d}{}z_{abcd}(\Phi_a^\dag\Phi_b)(\Phi_c^\dag\Phi_d)+...,
\end{equation*}
\begin{equation*}
V(\Phi_1,\Phi_2) ={} -\sum_{a,b}{}\mu_{ab}(\Phi_a^\dag\Phi_b) + \sum_{a,b,c,d}{}\lambda_{abcd}(\Phi_a^\dag\Phi_b)(\Phi_c^\dag\Phi_d).
\end{equation*}

    Thus one can always start with the action in the form (\ref{A2}) or (\ref{action}) without loss of generality.

\section{Higgs potential in the mass basis}
\label{apB}

The potential given in eq.~(\ref{genU}) can be written in terms of the mass eigenstates, which are massless Goldstone fields $G_0$, $G_+$, $G_-$ and massive Higgs bosons $h$, $H_0$, $A$, $H_+$, $H_-$,\footnote{For convenience, we rewrite $H^\pm$,  $G^0$, and $G^\pm$ as $H_\pm$,  $G_0$, and $G_\pm$, correspondingly.} in the following form
\begin{equation*}
V(h,H_0,A,H_\pm,G_0,G_\pm)=\frac{m_h^2}{2} h^2+\frac{m_H^2}{2} H_0^2+\frac{m_A^2}{2} A^2+m_{H^\pm}^2 H_+ H_- + I_3+I_4,
\end{equation*}
where
\begin{eqnarray*}
I_3 &=& \frac{v}{8} \left( g_p^2 \left\{ s_{\alpha+\beta} [c_{2 \alpha} h^3+c_{2 \beta}h(A^2-G_0^2-2G_- G_+)]
\right. \right.
\\
&+& c_{\alpha+\beta} [c_{2 \alpha} H_0^3-c_{2 \beta}H_0 (A^2-G_0^2-2 G_-G_+)] \nonumber \\
&+& \frac{h H_0}{2}[(c_{\alpha-\beta}-3 c_{3\alpha+\beta})h-(s_{\alpha-\beta}+3s_{3\alpha+\beta})H_0]\nonumber \\
&+& \left. 2 s_{2 \beta} A G_0(s_{\alpha+\beta} h-c_{\alpha+\beta}H_0) \right\}+
2 i g_2^2 A(H_+G_--H_-G_+) \nonumber\\
&+&h[(g^2s_{\alpha-\beta}+g_p^2s_{\alpha+3 \beta})H_+H_--(g_m^2c_{\alpha-\beta}+g_p^2 c_{\alpha+3 \beta})(H_+G_-+H_-G_+)] \nonumber\\
&-& \left. H_0[(g^2c_{\alpha-\beta}+g_p^2c_{\alpha+3 \beta})H_+H_-+(g_m^2s_{\alpha-\beta}+g_p^2 s_{\alpha+3 \beta})(H_+G_-+H_-G_+)] \right), \nonumber \\
\end{eqnarray*}
\begin{eqnarray*}
I_4 &=&\frac{g_p^2}{8} \left\{ -s_{4 \beta} [ AG_0(G_+G_--H_+H_-)+H_+G_+(G_-^2-H_-^2) \right. \\
&+&H_-G_-(G_+^2-H_+^2)+\frac{G_0^2-A^2}{2}(AG_0+H_+G_-+H_-G_+)] \nonumber \\
&-&2 c_{4 \beta}H_+H_-G_+G_-+s_{2 \beta}^2(G_+^2H_-^2+G_-^2H_+^2) \nonumber \\
&+& c_{2 \beta}^2 [ \frac{G_0^4+A^4}{4}+G_+G_-(G_+G_-+G_0^2)+H_+H_-(H_+H_-+A^2) ] \nonumber \\
&+&[c_{2 \beta}(A^2-G_0^2)+2s_{2 \beta}AG_0][c_{2 \alpha}\frac{(h^2-H_0^2)}{2}+s_{2 \alpha} hH_0]  \nonumber \\
&+&  \frac{1}{4} \left.\left[(1-3c_{4 \alpha})h^2 H_0^2+(1-3c_{4 \beta})A^2 G_0^2
+c_{2 \alpha}^2(h^4+H_0^4)+2s_{4 \alpha}hH_0(h^2-H_0^2)\right]\right\} \nonumber \\
&+& \mathrm{i} \frac{g_2^2}{4}(H_-G_+-H_+G_-)[s_{\alpha-\beta}(hA+H_0G_0)+c_{\alpha-\beta}(hG_0-H_0A)] \nonumber \\
&+& \frac{1}{4} \left[(g_1^2s_{2\beta}^2-g_2^2c_{2\beta}^2)AG_0(H_-G_++H_+G_-)-(g_1^2c_{2\beta}s_{2\alpha}+g_2^2s_{2 \beta}c_{2 \alpha})hH_0G_+G_-\right] \nonumber \\
&+& \frac{1}{16}[(2g_2^2+g_m^2c_{2(\alpha-\beta)}-g_p^2c_{2(\alpha+\beta)})(h^2G_+G_-+H_0^2H_+H_-) \nonumber \\
&+& (2g_2^2-g_m^2c_{2(\alpha-\beta)}+g_p^2c_{2(\alpha+\beta)})(H_0^2G_+G_-+h^2H_+H_-) \nonumber \\
&-& (g^2+g_p^2 c_{4 \beta})(H_+H_-G_0^2+G_+G_-A^2)] \nonumber \\
&+& \frac{1}{8} (g_1^2 s_{2 \beta} c_{2 \alpha}+g_2^2 c_{2 \beta} s_{2 \alpha})(h^2-H_0^2)(H_+G_-+H_-G_+) \nonumber \\
&+& \frac{hH_0}{8} \left[(g_p^2 s_{2(\alpha+\beta)}-g_m^2s_{2(\alpha-\beta)})H_+H_-
-(g_p^2c_{2(\alpha+\beta)}+g_m^2 c_{2(\alpha-\beta)})(H_-G_++H_+G_-)\right], \nonumber
\end{eqnarray*}
\begin{equation*}
m_h^2 = m_Z^2 s_{\alpha+\beta}^2+m_A^2 c_{\alpha-\beta}^2, \quad
m_H^2 = m_Z^2 c_{\alpha+\beta}^2+m_A^2 s_{\alpha-\beta}^2, \quad
m_{H_\pm}^2 = m_A^2+m_W^2,\quad m_W=\frac{v}{2} g_2,
\end{equation*}
$\sin \alpha=s_\alpha$, etc., and
\begin{equation*}
g_p^2=g_1^2+g_2^2, \qquad g_m^2=g_2^2-g_1^2, \qquad g^2=g_1^2-3g_2^2.
\end{equation*}

\end{document}